\documentclass{iopart}
\usepackage{iopams,graphicx,hyperref,accents,booktabs}
\usepackage[usenames,dvipsnames]{color}
\usepackage[normalem]{ulem}
\usepackage{subcaption, hhline}

\newcommand{\tfrac}[2]{\textstyle \frac{#1}{#2}}
\newcommand{\half}{\tfrac{1}{2}}
\newcommand{\third}{\tfrac{1}{3}}
\newcommand{\fourth}{\tfrac{1}{4}} 

\newcommand{\Lie}{\mathcal{L}}
\newcommand{\cM}{\mathcal{M}}
\newcommand{\cN}{\mathcal{N}}
\newcommand{\const}{\mathrm{const}}
\newcommand{\Kpp}{K^\phi{}_\phi}
\newcommand{\tf}{{\tilde f}}
\newcommand{\tU}{{\tilde U}}
\newcommand{\tW}{{\tilde W}}
\newcommand{\tX}{{\tilde X}} 
\newcommand{\tBp}{{\tilde B^\phi}}
\newcommand{\tE}{{\tilde E}}

\newcommand{\exps}{\rme^{2rs}}
\newcommand{\hp}{\hat p} 
\newcommand{\pn}{{\hat p}^n}
\newcommand{\mbe}{\mathbf{e}}
\newcommand{\mbc}{\mathbf{c}}
\newcommand{\mbm}{\mathbf{m}}
\newcommand{\mbs}{\mathbf{s}}
\newcommand{\mbfe}{\mathbf{f^e}}
\newcommand{\mbfm}{\mathbf{f^m}}
\newcommand{\mbke}{\mathbf{k^e}}
\newcommand{\mbkm}{\mathbf{k^m}}
\renewcommand{\Or}{\mathcal{O}} 

\newcommand{\Sectionref}[1]{section~\ref{#1}}

\newtheorem{proposition}{Proposition}

\begin{document}

\title{Dynamics of gravitational collapse in the axisymmetric Einstein-Vlasov system} 
\author{Ellery Ames$^{1}$, H\r{a}kan Andr\'easson$^{2}$ and Oliver Rinne$^{3}$}
\address{ 
  $^{1}$ Department of Mathematics, Humboldt State University, 1 Harpst St., Arcata, California 95521, USA\\ 
  $^{2}$ Mathematical Sciences, Chalmers University of Technology and University of Gothenburg, S-41296 Gothenburg, Sweden\\
  $^{3}$ Hochschule f\"ur Technik und Wirtschaft Berlin, Fachbereich 4, Treskowallee 8, 10318 Berlin, Germany}
\ead{ellery.ames@humboldt.edu, hand@chalmers.se, oliver.rinne@htw-berlin.de}
  
\begin{abstract}
  We numerically investigate the dynamics near black hole formation of solutions to the Einstein--Vlasov system in axisymmetry. 
  Our results are obtained using a particle-in-cell and finite difference code based on the $(2+1)+1$ formulation of the Einstein field equations in axisymmetry. 
  Solutions are launched from non-stationary initial data and exhibit type I critical behaviour. 
  In particular, we find lifetime scaling in solutions containing black holes, and support that the critical solutions are stationary.
  Our results contain examples of solutions that form black holes, perform damped oscillations, and appear to disperse. 
  We prove that complete dispersal of the solution implies that it has nonpositive binding energy.  
\end{abstract}
 


\section{Introduction}
\label{s:introduction}
Understanding the dynamics and qualitative features of self-gravitating matter, particularly beyond spherical symmetry and in the strong-field regime, remains a challenging area of study.  
In this paper we present numerical results on the dynamics and critical phenomena of general relativistic kinetic matter in the asymptotically flat and axisymmetric setting.
The assumption of axisymmetry reduces the computational complexity of the equations, and yet retains key features of the Einstein--matter system such as the presence of gravitational waves.

The kinetic matter model considered here represents a large collection of collisionless particle-like objects; so-called Vlasov matter. 
These particles move along geodesics in the spacetime generated by the entire collection.
We present spacetimes both with zero and non-zero net angular momentum. 
In the former case the particles have angular momentum even if the spacetime does not. 
The matter is modelled by a density function on the phase space of spacetime positions and momenta. 
In axisymmetry, the distribution function satisfies a transport equation in two spatial coordinates and three momentum coordinates. 
Our formulation of the Vlasov equation and the coupled Einstein--Vlasov system is presented in detail in section \ref{s:formulation}.

The Einstein--Vlasov system has been well studied in spherical symmetry, where global existence is proved for small initial data in the massive \cite{Rein1992} and massless \cite{Dafermos2006} cases. 
The formation of black holes from sufficiently strong ingoing data is also known \cite{Dafermos2005,Andreasson2010,AndreassonKR2011,Andreasson2012}.
In more general contexts, the stability of Minkowski space has been recently proved with both massive \cite{Fajman2017, Lindblad2019} and massless \cite{Taylor2017, Bigorgne2020} 
particles, and several spatially homogeneous solutions are shown to be future nonlinearly stable in \cite{RingstromBook2013}.
Here, we numerically study the dynamics and critical phenomena of axisymmetric solutions near the threshold of black hole formation.

Critical phenomena refer to the qualitative features displayed by solutions of the Einstein--matter equations (including the vacuum case) when initial data are tuned to the threshold of black hole formation. 
Typically, a one-parameter family of initial data is considered for which there exists a parameter value such that the evolved data collapse to a black hole, and another parameter value for which the solution launched by the data is free of singularities.
The critical data (the threshold for black hole formation) is that which divides these two end states. 
The numerical study of critical phenomena orginated in the work of Choptuik \cite{Choptuik1993} on the collapse of a massless scalar field in spherical symmetry. 
Critical behaviour exhibited by a model depends on the type of matter (if any) present. 
The massless scalar field, as well as several other field-theoretic models and fluids, display what is known as type II critical behaviour. 
Briefly, this means that the critical solution is self-similar and universal in the sense that it is independent of the initial data family. 
In addition, quantities such as the mass, angular momentum, and charge of black holes formed in the evolution of super-critical data are seen to scale as the parameter distance to the critical data with a universal exponent;
in particular, they become infinitesimally small at the black hole threshold.

In contrast, studies of critical collapse for Vlasov matter indicate type I critical behaviour. 
In type I, the critical solution is either an unstable stationary solution or time periodic.
Solutions launched from near-critical initial data evolve towards the critical solution, where they may linger for a time before either collapsing to a black hole or evolving towards a non-singular end state.
Scaling behaviour is observed in the duration of the evolution spent in the near-critical state, while the black hole mass and angular momentum remain finite and strictly bounded away from zero as the black hole threshold is approached from the supercritical side.

All investigations of critical behaviour in the Einstein--Vlasov system so far have been done in spherical symmetry.
In \cite{Rein1998}, Rein et al. find that for supercritical data, nearly all the mass of the initial data goes into the black hole, and only a small fraction of the mass escapes. 
They also observe a ``mass gap"; that is, a positive and finite value for the black hole mass at the black hole threshold.
Studies of Olabarrieta and Choptuik \cite{Olabarrieta2001} corroborate the existence of a mass gap, and also find support that the critical solutions in this setting are static spacetimes. 
Later work of Andr\'easson and Rein \cite{Andreasson2006} showed the static critical solutions cannot be universal. 
This observation is further supported in work of Akbarian and Choptuik \cite{Akbarian2014}, who, in addition, show that the exponent in the lifetime scaling of the intermediate state is not universal. 
The work of \cite{Andreasson2006} also suggests the existence of a periodic end state in addition to black hole formation and dispersal. 
This is confirmed in a recent detailed study by G\"{u}nther et al.~\cite{Gunther2020}, who numerically investigate the spherically symmetric Einstein--Vlasov system in three different coordinate systems. 
The authors find that perturbations of stable steady states oscillate, and in addition, that dispersal-promoting perturbations of unstable steady states also exhibit oscillatory behavior. 
Interestingly, there is evidence that such perturbations of unstable states, in cases, evolve towards, and oscillate about, an apparently stable steady state.

To our knowledge the present study is the first investigation of critical behaviour in the Einstein--Vlasov system beyond spherical symmetry. 
We present results of numerical simulations of solutions to the axisymmetric Einstein--Vlasov system both with and without net rotation. 
In particular, we find lifetime scaling in marginally supercritical solutions whose evolutions perform damped oscillations enroute to black hole formation.
We also find evidence that the critical states are non-unique stationary solutions. 
Similar to \cite{Gunther2020} we see cases in which a fraction of the particles are apparently expelled while the remnant recollapses, eventually approaching, via damped oscillations, an apparent steady state solution. 
The binding energy and rotation parameter $|J|/M^2$ (where $J$ and $M$ are the spacetime angular momentum and mass respectively)
are important in determining the possible end states of a given initial datum.
We prove that complete dispersion implies a non-positive binding energy.
Solutions with negative binding energy are observed to collapse to a black hole or disperse, but our results cannot rule out the recollapse and oscillations seen in \cite{Gunther2020}.

Critical collapse has been studied in axisymmetry for the vacuum Einstein system \cite{Abrahams1993,Garfinkle2001,Rinne2008,Hilditch2017}, radiation fluids \cite{Baumgarte2015,Baumgarte2016}, scalar fields \cite{Choptuik2003b,Choptuik2004}, and electromagnetic waves \cite{Baumgarte2019} (see \cite{Gundlach2007} for a detailed review and references).
All of these show type II critical behaviour.
Type I critical behaviour in axisymmetry has been found in boson and also neutron stars. 

Boson stars have also been observed to exhibit type I critical behaviour in the thesis of Lai \cite{Lai2004}. 
He studies the head-on collision of boson stars using the initial momentum parallel to the axis as the tuning parameter. 
Adding a self-interaction term in the model Lagrangian proved useful in being able to numerically distinguish sub and super solutions. 
The investigation is limited to what the author calls ``prompt" collapse, in which either a black hole is formed in the initial collision or it is not. 
Subcritical solutions (meaning ``promptly" subcritical solutions) are often observed to re-emerge as two stars and then subsequently collapse and form a black hole. 
Perturbations by a non-spherical real scalar field are also studied.
Support for a mass gap and lifetime scaling are given in both cases. 

Jin and Suen investigate in \cite{Jin2007} the head-on collision of two equal-mass neutron stars. 
Several different parameters are tuned to criticality and the authors observe lifetime scaling for the intermediate near critical state. 
The scaling exponent appears universal in that it is independent of the parameter varied, including parameters in the equation of state. 
Subsequent work by Kellerman, Radice, and Rezzolla \cite{Kellerman2010,Radice2010} verifies the time-periodic critical solution and lifetime scaling.
The authors argue however that in contrast to the conclusion of \cite{Jin2007,Wan2008}, the critical solution is in fact a perturbation of a linearly unstable spherically symmetric neutron star. 
We note similar behaviour is seen in magnetized neutron stars in \cite{Liebling2010}.
Subcritical solutions in this setting are observed to evolve towards neutron star configurations. 
In \cite{Radice2010} where the initial data consists of linearly unstable neutron stars, the authors provide evidence that the final state of subcritical data are stable neutron stars with the same mass as the initial data, up to small losses. 

Before proceeding we briefly discuss the numerical methods used in the present paper. 
We write the Einstein--matter equations using the $(2+1)+1$ formalism introduced by Maeda et al. \cite{Maeda1980}.
The result is a mixed system of elliptic and evolution equations for the metric fields coupled to the Vlasov equation. 
We solve this system with a second-order particle-in-cell (PIC) method using finite differences to solve for grid-based quantities.
Similar methods have been used in prior studies of the Einstein--Vlasov system. 
In particular, the results \cite{Rein1998,Olabarrieta2001,Andreasson2006,Gunther2020} in spherical symmetry mentioned above, all use PIC codes with finite differences. 
Such a PIC method is shown to converge in work of Rein and Rodewis \cite{ReinRodewis}.
We note that Akbarian and Choptuik in \cite{Akbarian2014} use finite-volume methods.
Particle methods have also been used beyond spherical symmetry to evolve the Einstein--Vlasov system. 
We mention early works by Shapiro and Teukolsky \cite{Shapiro1991,Shapiro1992a}; Abrahams et al. \cite{Abrahams1994}, who treat the case of net rotation; and recent work of East \cite{East2019}; all using axisymmetric codes. 
In addition, Shibata \cite{Shibata1999} and Yoo et al. \cite{Yoo2017} use fully 3D methods (though in \cite{Yoo2017} axisymmetric data are evolved).

This paper is organized as follows.
In section \ref{s:formulation.geometry} we present our formulation of the Einstein--matter equations in the $(2+1)+1$ formalism. 
In section \ref{s:formulation.matter} we specialize to Vlasov matter and derive the characteristic evolution equations. 
Our numerical methods are presented in section \ref{s:numerics}, and in section \ref{s:results} we present numerical results. 
Finally, in \ref{s:conclusion} we conclude with a discussion of our results. 
 

\section{Formulation of the Einstein-Vlasov system in axisymmetry}
\label{s:formulation}

In this section we derive the formulation of the Einstein-Vlasov
equations in axisymmetry that we set out to solve numerically.
In section \ref{s:formulation.geometry} we focus on the axisymmetric Einstein equations with general matter sources.
In section \ref{s:formulation.matter} we specialise the source terms to Vlasov matter, and we derive the equations of motion of that matter model.


\subsection{Geometry}
\label{s:formulation.geometry}


\subsubsection{(2+1)+1 reduction}
\label{s:formulation.geometry.reduction}

We assume that spacetime is axisymmetric with Killing vector $\xi$.
Cylindrical polar coordinates $t,r,z,\phi$ are chosen such that
$\xi = \partial/\partial\phi$.
We adopt the (2+1)+1 formalism \cite{Maeda1980}, which consists of a symmetry reduction w.r.t. the Killing vector followed by an Arnowitt-Deser-Misner (ADM) \cite{Arnowitt1962, York1979} reduction of the symmetry-reduced spacetime.

In the first step, spacetime $(\cM, g_{\alpha\beta})$ is reduced to a three-dimensional Lorentzian spacetime $\cN$ formed by the trajectories of the Killing vector \cite{Geroch1971}.
The information contained in the spacetime metric $g_{\alpha\beta}$ is split up into the norm $\lambda$ of the Killing vector,
\begin{equation}
  \lambda^2 = g_{\alpha\beta} \xi^\alpha \xi^\beta,
\end{equation}
the metric $h_{\alpha\beta}$ on $\cN$,
\begin{equation}
  \label{e:geroch_h}
  h_{\alpha\beta} = g_{\alpha\beta} - \lambda^{-2} \xi_\alpha \xi_\beta,
\end{equation}
and the \emph{twist vector}
\begin{equation}
  \label{e:twistvector}
  \omega_\alpha =
  \varepsilon_{\alpha\beta\gamma\delta} \xi^\beta \nabla^\gamma \xi^\delta ,
\end{equation}
where $\varepsilon_{\alpha\beta\gamma\delta}$ is the alternating symbol of $g_{\alpha\beta}$ and $\nabla$ the metric connection of $g_{\alpha\beta}$.
We also define the alternating symbol of $h_{\alpha\beta}$,
\begin{equation}
  \label{e:eps3}
  \varepsilon_{\alpha\beta\gamma} = \lambda^{-1} \varepsilon_{\alpha\beta\gamma\delta} \xi^\delta.
\end{equation}
The energy-momentum tensor is decomposed as
\begin{equation}
\label{eq:EMtensorN}
  \tau = \lambda^{-2} \xi^\gamma \xi^\delta T_{\gamma\delta}, \quad
  \tau_\alpha = \lambda^{-2} h_\alpha{}^\gamma \xi^\delta T_{\gamma\delta}, \quad
  \tau_{\alpha\beta} = h_\alpha{}^\gamma h_\beta{}^\delta T_{\gamma\delta} .
\end{equation}
The Lie derivatives w.r.t.~$\xi$ and the contractions with $\xi$ of the
tensor fields $\lambda, h_{\alpha\beta}, \varepsilon_{\alpha\beta\gamma}, \omega_\alpha,
\tau, \tau_{\alpha}, \tau_{\alpha\beta}$ vanish.
Thus we write them with indices $a,b,\ldots$ ranging over $t,r,z$ in the
following.

In the second step, an ADM reduction \cite{Arnowitt1962, York1979} is applied to $\cN$, i.e.~it is foliated by spacelike hypersurfaces $\Sigma(t)$ with future-directed unit normal $n_a$ and induced metric
\begin{equation}
  H_{ab} = h_{ab} + n_a n_b.
\end{equation}
Thus, $h_{ab}$ is decomposed as
\begin{equation}
\label{e.HmetricDefinition}
  h = -\alpha^2 \rmd t^2
    + H_{AB} (\rmd x^A + \beta^A \rmd t)(\rmd x^B + \beta^B \rmd t),
\end{equation}
where $\alpha$ is the lapse function and $\beta^A$ the shift vector, and
indices $A,B, \ldots$ range over $r, z$.

The extrinsic curvature of the $\Sigma(t)$ slices is given by
\begin{equation}
  \label{e:chidef}
  \chi_{ab} = -\half \Lie_n H_{ab}.
\end{equation}
We also introduce the $\phi\phi$-component of the extrinsic curvature,
\begin{equation}
  \label{e:Kpp}
  K^\phi{}_\phi = -\lambda^{-1} n^a \lambda_{,a}.
\end{equation}
The twist vector is decomposed as
\begin{equation}
  \label{e:ADMtwist}
  E^a = \lambda^{-3} \varepsilon^{ab} \omega_b, \qquad
  B^\phi = \lambda^{-3} n^a \omega_a,
\end{equation}
where $\varepsilon_{ab} = n^c \varepsilon_{cab}$ is the alternating symbol of $H$.
Finally the energy-momentum tensor is further decomposed as
\begin{eqnarray} \label{e:mattersources}
  J^\phi = -n_a \tau^a , \quad
  S_A = H_{Aa} \tau^a , \quad
  \rho_H = n^a n^b \tau_{ab}, \nonumber\\
  J_A = - H_A{}^a n^b \tau_{ab}, \quad
  S_{AB} = H_A{}^a H_B{}^b \tau_{ab} .
\end{eqnarray}

The general forms of the Einstein equations and energy-momentum conservation equations using these definitions of the reduced variables can be found in \cite{Rinne2005}.


\subsubsection{Choice of gauge and variables}

We choose maximal slicing $\chi^A{}_A + \Kpp = 0$.
Preservation of this condition under the time evolution implies
\begin{eqnarray} \label{e:maxsligen}
  \fl \alpha_{\parallel A}{}^A
  - \alpha \big[ \chi^{AB} \chi_{AB} + (\Kpp)^2
  - \lambda^{-1} \alpha^{-1} \lambda_{,A} \alpha^{,A}
  + \half \lambda^2 E_A E^A \nonumber\\
  + \half \kappa \left( \rho_H + \tau + S_A{}^A \right) \big] = 0,
\end{eqnarray}
where $\parallel$ denotes the covariant derivative compatible with $H$, and $\kappa = 8\pi$ in geometric units.

In quasi-isotropic (or isothermal) gauge, we may write the
2-metric and norm of the Killing vector as \cite{Abrahams1993,Garfinkle2001,Choptuik2003a,Rinne2008}
\begin{equation} \label{e:quasi-isotropic}
  H = \psi^4 \rme^{2rs} (\rmd r^2 + \rmd z^2), \qquad
  \lambda = r \psi^2.
\end{equation}
The relevant regularity condition relating the $rr$ and $\phi\phi$ components of smooth axisymmetric tensor fields \cite{Rinne2005} is satisfied provided that $s = \Or(r)$ on the axis.
Preservation of \ref{e:quasi-isotropic} under the time evolution implies the Cauchy-Riemann-like equations
\begin{eqnarray}
    \label{e:shiftr}
    0 = \beta^r{}_{,r} - \beta^z{}_{,z} + \alpha (\chi^z{}_z - \chi^r{}_r), \\
    \label{e:shiftz}
    0 = \beta^r{}_{,z} + \beta^z{}_{,r} - 2 \alpha \chi^r{}_z.
\end{eqnarray}  

We define the following combinations of extrinsic curvature components,
\begin{eqnarray}
  \tW := r^{-1} \psi^6 (\chi^r{}_r - K^\phi{}_\phi), \quad
  \tU := \psi^6 (\chi^z{}_z - \chi^r{}_r), \quad
  \tX := \psi^6 \chi^r{}_z.
\end{eqnarray}
The factor of $r^{-1}$ in the definition of $\tW$ ensures that this quantity is $\Or(r)$ on the axis (\ref{s:appendix.geodetails.bcs}).
The rescaling with $\psi$ has been chosen such that $\psi$ drops out of the extrinsic curvature terms in the linear momentum constraints \eref{e:momconsr} and \eref{e:momconsz} below, and such that the Hamiltonian constraint \eref{e:hamcons} becomes a Helmholtz-like equation with the ``right sign'' so that potential problems with non-uniqueness of solutions are avoided \cite{Rinne2008,Walsh2007}.

The twist variables are rescaled for the same reason:
\begin{equation}
  \label{eq:ScaledTwists}
  \tE^A := \psi^7 E^A, \quad \tBp := \psi^{7} B^\phi.
\end{equation}  

The matter sources are rescaled such that they become independent of the metric fields, cf. \eref{e:tautildeframe}--\eref{e:Sddtildeframe}:
\begin{eqnarray}
\label{eq:ScaledMatter}	
\tilde \tau 		:= \psi^6 e^{2rs} \tau , \quad
\tilde J^{\phi} 	:= \psi^8 e^{2rs} J^{\phi} , \quad
\tilde S_{A} 	:= \psi^{6} e^{rs} S_{A} , \\
\tilde \rho_H 	:= \psi^{6} e^{2rs} \rho_H , \quad
\tilde J_A 		:= \psi^{4} e^{rs} J_A , \quad
\tilde S_{AB} 	:= \psi^{2} S_{AB} .\nonumber
\end{eqnarray}

The Einstein equations split into time-independent elliptic constraint equations and time-dependent evolution equations. 
Our gauge conditions are also of elliptic type.
In the following, we write out explicitly all the elliptic and evolution equations that we solve numerically. 
We use the abbreviations
\begin{equation}
  P_A := \psi^{-1} \psi_{,A}, \quad R_A := (rs)_{,A}, \quad
  A_A := \alpha^{-1} \alpha_{,A}.
\end{equation}


\subsubsection{Elliptic equations}

The linear momentum constraints form a first-order elliptic system for
$\tU$ and $\tX$:
\begin{eqnarray}
  \label{e:momconsr}
  \fl 0 = 2 \tX_{,z} - \tfrac{2}{3} \tU_{,r} + \tfrac{2}{3} r \tW_{,r}
    + 4 R_z \tX - 2 R_r \tU + \tfrac{8}{3} \tW - r^2 \exps \tBp \tE^z 
    \nonumber\\
    - 2 \kappa e^{-rs} \psi^2 \tilde J_r, \\
  \label{e:momconsz}
  \fl 0 = 2 \tX_{,r} + \tfrac{4}{3} \tU_{,z} + \tfrac{2}{3} r \tW_{,z}
    + (4 R_r + 2 r^{-1}) \tX + 2 R_z \tU + r^2 \exps \tBp \tE^r
    \nonumber\\
    - 2 \kappa e^{-rs} \psi^2 \tilde J_z.
\end{eqnarray}
The angular momentum constraint is an equation for the divergence of $\tE^A$:
\begin{equation}
  \label{e:gercons}
  \fl 0 = \tE^r{}_{,r} + \tE^z{}_{,z} + (3 P_r + 2 R_r + 3 r^{-1}) \tE^r
    + (3 P_z + 2 R_z) \tE^z - 2 \kappa e^{-2rs} \psi^{-1} \tilde J^\phi.
\end{equation}
The Hamiltonian constraint is an elliptic equation for $\psi$:
\begin{eqnarray}
  \label{e:hamcons}
  \fl 0 = \psi_{,rr} + \psi_{,zz} + r^{-1} \psi_{,r} + \fourth \psi [ r s_{,rr} +
    r s_{,zz} + 2 s_{,r} ] \nonumber\\ + \fourth \psi^{-7} \exps \left[
    \third (\tU + \half r \tW)^2 + \fourth (r \tW)^2 + \tX^2 \right]
  \nonumber\\
  + \tfrac{1}{16} r^2 \psi^{-5} \exps \left\{ (\tBp)^2 + \psi^4 \exps
    \left[(\tE^r)^2 + (\tE^z)^2\right]\right\} \nonumber\\
  + \fourth \kappa \psi^{-1} \tilde \rho_H .
\end{eqnarray}
The maximal slicing condition \eref{e:maxsligen} is an elliptic equation
for $\alpha$:
\begin{eqnarray}
  \label{e:slicing}
  \fl 0 = \alpha_{,rr} + \alpha_{,zz} + (2 P_r + r^{-1}) \alpha_{,r}
  + 2 P_z \alpha_{,z} \nonumber\\
  - 2\alpha \psi^{-8} \rme^{2rs} \left[ \third (\tU + \half r \tW)^2
    + \fourth (r \tW)^2 + \tX^2 \right] \nonumber\\
  - \half r^2 \alpha \psi^{-2} \rme^{4rs} \left[(\tE^r)^2 + (\tE^z)^2\right]
  \nonumber\\
  - \half \kappa \alpha \psi^{-2} \left[ \tilde \rho_H + \tilde \tau
  + \tilde S_{rr} + \tilde S_{zz} \right].
\end{eqnarray}
In order to write the quasi-isotropic gauge conditions \eref{e:shiftr} and \eref{e:shiftz} in a form that is better suited to numerical solution, we express the shift in terms of two potentials $P$ and $Q$,
similarly to \cite[Chapter IV e)]{EvansPhD}:
\begin{equation} \label{e:shiftpotdef}
  \beta^r = P_{,r} + Q_{,z}, \qquad 
  \beta^z = Q_{,r} - P_{,z}.
\end{equation}  
This turns \eref{e:shiftr} and \eref{e:shiftz} into two decoupled Poisson equations:  
\begin{eqnarray}
  \label{e:shiftpot1}
  0 = P_{,rr} + P_{,zz} + \alpha \psi^{-6} \tU, \\
  \label{e:shiftpot2}
  0 = Q_{,rr} + Q_{,zz} - 2 \alpha \psi^{-6} \tX.
\end{eqnarray}
Once we have solved for $P$ and $Q$, the shift is formed from \eref{e:shiftpotdef}.

We impose boundary conditions on the elliptic equations at the outer boundary of the computational domain compatible with asymptotic flatness as discussed in \ref{s:appendix.geodetails.bcs}.


\subsubsection{Evolution equations}

The evolution equations are
\begin{eqnarray}
  \label{e:dtpsi}
  \fl \psi_{,t} = \beta^r \psi_{,r} + \beta^z \psi_{,z} + \half r^{-1} \beta^r \psi + \tfrac{1}{6} \alpha \psi^{-5} (\tU + 2 r \tW), \\
  \label{e:dts}
  \fl s_{,t} = \beta^r s_{,r} + \beta^z s_{,z} - \alpha \psi^{-6} \tW + r^{-1} \beta^r s + (r^{-1} \beta^r)_{,r} , \\
  \label{e:dtW_}
  \fl \tW_{,t} = \beta^r \tW_{,r} + \beta^z \tW_{,z} + 4 r^{-1} \beta^r \tW + 2 r^{-1} \tX \beta^z{}_{,r} \nonumber\\
  - \alpha \psi^2 \rme^{-2rs} \left[ \alpha^{-1} (r^{-1} \alpha_{,r})_{,r} + 2 \psi^{-1} (r^{-1} \psi_{,r})_{,r} + s_{,rr} + s_{,zz} \right.\nonumber\\ \qquad
  + (r^{-1} s)_{,r} - r^{-1} R_r (A_r + 2 P_r) + r^{-1} R_z (A_z + 2 P_z) \nonumber\\
  \qquad \left. - 2 r^{-1} P_r (2 A_r + 3 P_r) \right] \nonumber\\
  - \half \alpha \psi^{-6} r^{-1} \left[ 4 \tX^2 - 2 r\tW\tU - 4 (r\tW)^2 \right]\nonumber\\
  + \alpha r \left\{ \exps \left[(\tE^r)^2 + \half (\tE^z)^2\right] 
  - \psi^{-4} (\tBp)^2 \right\}\nonumber\\
  + \kappa \alpha r^{-1} e^{-2rs} (\tilde \tau - \tilde S_{rr}),\\
  \label{e:dtU_}
  \fl \tU_{,t} = \beta^r \tU_{,r} + \beta^z \tU_{,z} - 4 \beta^z{}_{,r} \tX + 3 r^{-1} \beta^r \tU \nonumber\\
  + \alpha \psi^2 \rme^{-2rs} \left[ \alpha^{-1} (\alpha_{,rr} - \alpha_{,zz}) + 2 \psi^{-1} (\psi_{,rr} - \psi_{,zz}) \right.\nonumber\\
  \qquad - 2 A_r (2 P_r + R_r) + 2 A_z (2 P_z + R_z) - 2 P_r (3 P_r + 2 R_r) \nonumber\\ 
  \qquad \left. + 2 P_z (3 P_z + 2 R_z) - 2 r^{-1} R_r \right]
  \nonumber\\
  + \alpha \psi^{-6} \left[ \tU(\tU + 2 r \tW) + 4 \tX^2 \right] + \half r^2 \alpha \exps \left[ (\tE^z)^2 - (\tE^r)^2 \right]\nonumber\\
  + \kappa \alpha \rme^{-2rs} (\tilde S_{rr} - \tilde S_{zz}),\\
  \label{e:dtX_}
  \fl \tX_{,t} = \beta^r \tX_{,r} + \beta^z \tX_{,z} + 3 r^{-1} \beta^r \tX + \beta^z{}_{,r} \tU \nonumber\\
  + \alpha \psi^2 \rme^{-2rs} \left[ -\alpha^{-1} \alpha_{,rz} - 2\psi^{-1} \psi_{,rz} + A_r(R_z + 2P_z) \right.\nonumber\\\left.
  \qquad + A_z (R_r + 2 P_r) + 2 P_r R_z + 2 P_z R_r + 6 P_r P_z + s_{,z} \right] \nonumber\\
  + 2 \alpha \psi^{-6} r\tW \tX + \half r^2 \alpha \exps \tE^r \tE^z - \kappa \alpha \rme^{-2rs} \tilde S_{rz},\\
  \label{e:dtEr_}
  \fl \tE^r{}_{,t} = \beta^r \tE^r{}_{,r} + \beta^z \tE^r{}_{,z}
    + \alpha \psi^{-4} \rme^{-2rs} \left[\tBp{}_{,z} + (A_z - P_z) \tBp
        \right]\nonumber\\
    + \tE^r \left[ -\beta^r{}_{,r} + \tfrac{7}{2} r^{-1} \beta^r
    + \half \alpha \psi^{-6} (2r\tW + \tU) \right] - \tE^z \beta^r{}_{,z}
    \nonumber\\
    - 2 \kappa\alpha \psi^{-3} \rme^{-3rs} \tilde S_r,\\
  \label{e:dtEz_}
  \fl \tE^z{}_{,t} = \beta^r \tE^z{}_{,r} + \beta^z \tE^z{}_{,z}
    - \alpha \psi^{-4} \rme^{-2rs} \left[\tBp{}_{,r} + (A_r - P_r + 3 r^{-1}) \tBp
      \right]\nonumber\\
    + \tE^z \left[ -\beta^z{}_{,z} + \tfrac{7}{2} r^{-1} \beta^r
    + \half \alpha \psi^{-6} (2r\tW + \tU) \right] - \tE^r \beta^z{}_{,r}
    \nonumber\\
    - 2 \kappa\alpha \psi^{-3} \rme^{-3rs} \tilde S_z,\\
  \label{e:dtBp_}
  \fl \tBp{}_{,t} = \beta^r \tBp{}_{,r} + \beta^z \tBp{}_{,z}
    + \tBp \left[\tfrac{7}{2} r^{-1} \beta^r
      + \tfrac{3}{2} \alpha \psi^{-6} (2r\tW + \tU)\right]
    \nonumber\\
    + \alpha \left[ \tE^r{}_{,z} - \tE^z{}_{,r} + \tE^r(A_z - 3 P_z + 2 R_z)
    - \tE^z (A_r - 3 P_r + 2 R_r) \right].
\end{eqnarray}

We impose boundary conditions on the evolution equations at the outer boundary of the computational domain compatible with asymptotic flatness as discussed in \ref{s:appendix.geodetails.bcs}.


\subsubsection{Evolution scheme}
\label{s:formulation.geometry.scheme}

We choose to solve the Hamiltonian constraint \eref{e:hamcons} for $\psi$ but evolve the extrinsic curvature variables $\tU$ and $\tX$ using their evolution equations \eref{e:dtU_} and \eref{e:dtX_}.
The redundant equations, namely the evolution equation \eref{e:dtpsi} for $\psi$ and the momentum constraints \eref{e:momconsr} and \eref{e:momconsz} for $\tU$ and $\tX$, are used to monitor the accuracy of the code.

To summarise, the evolution equations \eref{e:dts}--\eref{e:dtBp_} are used to evolve the variables $s, \tW, \tU, \tX, \tE^r, \tE^z, \tBp$.
At each time step, the Hamiltonian constraint \eref{e:hamcons} is solved for $\psi$, the maximal slicing condition \eref{e:slicing} is solved for $\alpha$, and (the potential version of) the isotropic gauge conditions
\eref{e:shiftpotdef}--\eref{e:shiftpot2}  are solved for $\beta^r$ and $\beta^z$. 
The methods we use for solving these equations numerically are presented in section \ref{s:numerics}.

The linear and angular momentum constraints \eref{e:momconsr}, \eref{e:momconsz} and \eref{e:gercons} still need to be solved to construct the initial data.
How this is done is explained in \ref{s:appendix.geodetails.iniconstr}.
Boundary conditions for the elliptic and evolution equations are discussed in \ref{s:appendix.geodetails.bcs}.


\subsection{Vlasov matter}
\label{s:formulation.matter}


\subsubsection{Energy-momentum tensor}

Vlasov matter is described by a distribution function $f(x^\alpha, p_\alpha)$ of the spacetime coordinates $x^\alpha$ and momentum $p_\alpha$. 
(We prefer to use the covariant components $p_\alpha$ of the momentum in the following; one can also regard $f$ as a function of the contravariant components $p^\alpha$.)
The energy-momentum tensor on the cotangent bundle is given by 
\begin{equation}\label{e:emcotan}
  T_{\alpha\beta}=\int f\,\frac{p_{\alpha}p_{\beta}}{m}\frac{1}{\sqrt{|g|}}\rmd p_0\rmd p_1\rmd p_2\rmd p_3.
\end{equation}
By using the mass shell condition
\begin{equation}
  \label{e:massshell}
  g_{\alpha\beta}p^{\alpha}p^{\beta}=-m^2,
\end{equation}
where $m$ is the rest mass of the particles, 
and the assumption that all particles have the same rest mass, we obtain
\begin{equation}\label{e:emcotan2}
  T_{\alpha\beta}=\int f\,\frac{p_{\alpha}p_{\beta}}{p^0}\frac{1}{\sqrt{|g|}}\rmd p_1 \rmd p_2\rmd p_3.
\end{equation}
In what follows we set $m=1$.

Following the $(2+1)+1$ formalism of section \ref{s:formulation.geometry}, we first decompose the momentum $p_\alpha$ w.r.t. the Killing vector $\xi$:
\begin{equation}
  L := \xi^\alpha p_\alpha, \qquad
	\hat p_b := h_b{}^\alpha p_\alpha,
\end{equation}  
and then we further decompose $\hat p_a$ w.r.t. the timelike normal $n$:
\begin{equation}
  \hat p^n := n^a \hat p_a, \qquad
  q_A  : = H_A{}^b \hat p_b.
\end{equation}  
Again, indices $\alpha,\beta,\ldots$ range over $t,r,z,\phi$, indices $a,b,\ldots$ over $t,r,z$, and indices $A,B,\ldots$ over $r,z$.

Using the mass shell condition \eref{e:massshell}, the component $p^0 = \hat p^0 = -\alpha^{-1} \hat p^n$ in \eref{e:emcotan2} is given by
\begin{equation}
p^0 = \alpha^{-1} \sqrt{H^{AB} q_A q_B + \lambda^{-2} L^2 + 1}.
\end{equation}

With these definitions the matter source terms \eref{eq:EMtensorN}, \eref{e:mattersources} take the form 
\begin{eqnarray}
  \tau = \lambda^{-2} \frac{\alpha}{\sqrt{|g|}} \int f(q, L) \frac{L^2 \, \rmd q_1\rmd q_2\rmd L }{\sqrt{H^{AB} q_A q_B + \lambda^{-2} L^2 + 1}},\\
  J^\phi	= \lambda^{-2} \frac{\alpha}{\sqrt{|g|}} \int f(q, L) L \,\rmd q_1\rmd q_2\rmd L,\\
  S_A =  \lambda^{-2} \frac{\alpha}{\sqrt{|g|}} \int f(q, L) \frac{q_A L \, \rmd q_1\rmd q_2\rmd L }{\sqrt{H^{AB} q_A q_B + \lambda^{-2} L^2 + 1}},\\
  \rho_H 	= \frac{\alpha}{\sqrt{|g|}} \int f(q, L)  \sqrt{H^{AB} q_A q_B + \lambda^{-2} L^2 + 1} \   \rmd q_1\rmd q_2\rmd L,\\
  J_A = \frac{\alpha}{\sqrt{|g|}} \int f(q, L) q_A  \rmd q_1\rmd q_2\rmd L ,\\
  S_{AB} = \frac{\alpha}{\sqrt{|g|}} \int f(q, L)\frac{ q_A q_B\, \rmd q_1\rmd q_2\rmd L}{\sqrt{H^{AB} q_A q_B + \lambda^{-2} L^2 + 1}}.
\end{eqnarray}
  
It is convenient to introduce the following frame in the momentum variables
\begin{equation}
\label{e:MomentumFrame}
  v_1 = e^{-rs} \psi^{-2} q_1, \quad v_2 = e^{-rs} \psi^{-2} q_2, \quad
  v_3 = r^{-1} \psi^{-2} L.
\end{equation}
In terms of this frame we have
\begin{equation} 
    \alpha p^0 = \sqrt{H^{AB} q_A q_B + \lambda^{-2} L^2 + 1} = \sqrt{v_1^2 + v_2^2 + v_3^2 + 1}
\end{equation}
and
\begin{equation} 
    \rmd q_1 \rmd q_2 \rmd L  = r e^{2rs} \psi^{6} \rmd v_1 \rmd v_2 \rmd v_3 . 
\end{equation}

If we also rescale the Vlasov distribution function,
\begin{equation}
  \label{eq:RescaledDensity}
  \tilde f := e^{2rs} \psi^6 f,
\end{equation}
then the rescaled matter source terms \eref{eq:ScaledMatter} become independent of the metric:
\begin{eqnarray}
  \label{e:tautildeframe}
  \tilde \tau 		= & \int \frac{\tilde f(v)  v_3^2 \,\rmd v_1\rmd v_2\rmd v_3}{\sqrt{v_1^2 + v_2^2 + v_3^2 + 1}}, \\
  \label{e:Jphitildeframe}
  \tilde J^\phi 	= & r^{-1} \int \tilde f(v) v_3 \, \rmd v_1 \rmd v_2 \rmd v_3 ,\\
  \label{e:Sdtildeframe}
  \tilde S_A 		= & r^{-1} \int \frac{\tilde f(v)  v_A v_3 \, \rmd v_1 \rmd v_2 \rmd v_3 }{ \sqrt{v_1^2 + v_2^2 + v_3^2 + 1} }, \\
  \tilde \rho_H 	= & \int \tilde f(v) \sqrt{v_1^2 + v_2^2 + v_3^2 + 1} \,\rmd v_1\rmd v_2\rmd v_3,  \\
  \tilde J_A 		= & \int \tilde f(v) v_A \,\rmd v_1\rmd v_2\rmd v_3,  \\
  \label{e:Sddtildeframe}
  \tilde S_{AB} 	= &  \int \tilde f(v) v_A v_B\frac{ \,\rmd v_1\rmd v_2\rmd v_3}{\sqrt{v_1^2 + v_2^2 + v_3^2 + 1}} .
\end{eqnarray}


\subsubsection{Geodesic equation}
\label{s:formulation.matter.geodesic_eqns}
Particles in the Vlasov model move along geodesics, which are given by the geodesic equation
\begin{equation}
  p^\beta \nabla_\beta p^\alpha = 0.
\end{equation}

First decomposing this equation w.r.t. the Killing vector, we obtain
\begin{eqnarray}
  0 = \hp^b D_b L, \\
  \label{e:geodesic_geroch}
  0 = \hp^b D_b \hp^a + \lambda^{-3} L \varepsilon^{abc} \hp_b \omega_c - \lambda^{-3} L^2 \lambda^a,
\end{eqnarray}
where $D$ is the covariant derivative of the metric $h$ on $\mathcal{N}$.
The first equation tells us that angular momentum is conserved.

Next, we perform an ADM decomposition of \eref{e:geodesic_geroch}.
After a lengthy calculation we arrive at 
\begin{eqnarray}
  \fl \Lie_n \pn = (\pn)^{-1} \left[ q^B d_B \pn + L E^B q_B + L^2 \lambda^{-2} \Kpp \right], \\
  \label{e:LienqA}
  \fl \Lie_n q_A = (\pn)^{-1} \left[ q^B d_B q_A + (\pn)^2 \alpha^{-1} \alpha_{,A }- L \varepsilon_{AB} q^B B^\phi + L \pn E_A - L^2 \lambda^{-3} \lambda_{,A} \right],
\end{eqnarray}
where $d$ is the covariant derivative of the spatial 2-metric $H$ and $\Lie_n = \alpha^{-1}(\partial_t - \Lie_\beta)$ is the Lie derivative along the unit timelike normal.

We need to write the geodesic equation as a first-order system of ordinary differential equations (ODEs) w.r.t.~coordinate time $t$.
Letting $\sigma$ be an affine parameter along the geodesic,
\begin{eqnarray}
  \frac{\rmd x^A}{\rmd t} = \frac{\rmd \sigma}{\rmd t} \frac{\rmd x^A}{\rmd \sigma} = -\alpha (\pn)^{-1} H^{AB} q_B - \beta^A,\\
  \frac{\rmd q_A}{\rmd t} = \frac{\rmd x^\beta}{\rmd t} \frac{\partial q_A}{\partial x^\beta} = \frac{\partial q_A}{\partial t} - \left[ \alpha (\pn)^{-1} H^{BC} q_C + \beta^B \right] \frac{\partial q_A}{\partial x^B},
\end{eqnarray}
and $\partial q_A/\partial t$ can be inferred from \eref{e:LienqA}.

In terms of our frame variables \eref{e:MomentumFrame}, these equations take the explicit form 
\begin{eqnarray}
\label{e:drdt}
\fl \frac{\rmd r}{\rmd t} = -\beta^1 +  \frac{\psi^{-2}  e^{-r s}}{p^0} v_1,\\
\label{e:dzdt}
\fl \frac{\rmd z}{\rmd t} = -\beta^2 +\frac{ \psi^{-2}  e^{-r s}}{p^0} v_2,\\
\label{e:dv1dt}
\fl \frac{\rmd v_1}{\rmd t} = -\psi^{-2}e^{-rs} \alpha \alpha_{,r} p^0 - \frac 13 \psi^{-6}\alpha ( \tilde U - r \tilde W)v_1 + v_2 \beta^2_{,r} - \frac{\psi^{-2}e^{-rs}}{p^0} \left( R_z + 2P_z\right) v_1 v_2 	\nonumber\\
+ \frac{\psi^{-2}e^{-rs}}{p^0} \left( R_r + 2P_r\right) v_2^2 + \frac{\psi^{-2}e^{-rs}}{p^0} \left( r^{-1} + 2P_r\right) v_3^2 \nonumber\\
+ \alpha e^{rs} \psi^{-3} v_3 r\tE^r + \psi^{-5} (p^0)^{-1} v_2v_3 r\tBp, \\
\label{e:dv2dt}
\fl \frac{\rmd v_2}{\rmd t} = -\psi^{-2}e^{-rs} \alpha \alpha_{,z} p^0 - \frac{\psi^{-6} \alpha}{3}(\tilde U - r \tilde W) v_2 + v_1 \beta^1_{,z}  + v_2 ( \beta^2_{,z}- \beta^1_{,r} ) \nonumber\\
+ \frac{\psi^{-2}e^{-rs}}{p^0} \left( R_z + 2 P_z\right) v_1^2 - \frac{\psi^{-2}e^{-rs}}{p^0} \left( R_r + 2 P_r\right) v_1v_2 \nonumber\\
+ 2 \frac{\psi^{-2}e^{-rs}}{p^0} P_z v_3^2 + \alpha e^{rs} \psi^{-3} v_3 r\tE^z - \psi^{-5} (p^0)^{-1} v_1v_3 r\tBp, \\
\label{e:dv3dt}
\fl \frac{\rmd v_3}{\rmd t} = - \frac 13 \psi^{-6} \alpha (\tilde U + 2 r \tilde W) v_3 - \frac{\psi^{-2}  e^{-r s}}{rp^0} v_1v_3 - \frac{2\psi^{-2}e^{-rs}}{p^0}\left( P_r v_1 + P_z v_2 \right) v_3. 
\end{eqnarray}
Note that since $v_3 = r^{-1} \psi^{-2} L$, this momentum is not preserved under the evolution, even though $L$ is.
We note, in view of the characteristic system above, that the Vlasov equation for the density function $f$ reads
\begin{equation}
  f_{,t}+\frac{\rmd r}{\rmd t}f_{,r}+\frac{\rmd z}{\rmd t}f_{,z}+\frac{\rmd v_1}{\rmd t}f_{,v_1}+\frac{\rmd v_2}{\rmd t}f_{,v_2}+\frac{\rmd v_3}{\rmd t}f_{,v_3}=0.
\end{equation}
When we introduce the frame \eref{e:MomentumFrame}, the geodesic flow is no longer volume preserving due to the choice of non-canonical momentum variables. 
The conserved quantity is 
\begin{equation}
  \int \int e^{2rs}\psi^6 r f \rmd x \,  \rmd v = \int \int r \tilde f \rmd x \, \rmd v,
\end{equation}
and it follows that $\tilde{f}$ obeys the evolution equation (cf.~Lemma 2.1 in \cite{ReinRodewis}) 
\begin{equation}
  \label{e:dfdt}
  \frac{\rmd \tilde f}{\rmd t} = \tilde f \left[r^{-1}\beta^r 
  - \psi^{-2}e^{-rs}(p^0)^{-1}r^{-1}v_1 \right]. 
\end{equation}
Here the derivative is taken along a characteristic. 
 
\section{Numerical methods}
\label{s:numerics}

The system described above consists of a nonlinear set of partial differential equations in two spatial dimensions plus time for the metric fields. 
This system is coupled to the matter quantities and is solved by a particle-in-cell (PIC) method \cite{HockneyEastwood1988}. Particle-based methods have been used in numerical relativity by several authors, including early work of Shapiro, Teukolsky and collaborators \cite{Shapiro1985a,Shapiro1985b,Rasio1989,Shapiro1992a,Shapiro1992b,Abrahams1994}, as well as more recent work by others \cite{Olabarrieta2001,Andreasson2006,Yoo2017,East2019}.

Our numerical method is presented below. Most routines have been written from scratch in \textsc{Python} and \textsc{C}. 
We make use of core functionality of these languages as well as the \textsc{NumPy} and \textsc{SciPy} libraries, and the \textsc{Umfpack} linear solver. 
The code is parallelized using the \textsc{OMP} library on a shared memory architecture. \footnote{\url{https://numpy.org/}, \url{https://www.scipy.org/}, \url{https://pypi.org/project/scikit-umfpack/}, \url{http://www.openmp.org/}}

\subsection{Computational domain and grid}
The numerical domain is a subset of a meridional plane $(r,z) \in [0, r_{\max}] \times [-z_{\max}, z_{\max}]$. A rectilinear grid is formed by placing $N_r + 1$ grid points in the $r-$direction at 
\[ r_i = \left(\frac 12 + i\right) h_r, \quad i = 0, \ldots N_r, \quad h_r = \frac{r_{\max}}{N_r + \frac 12},\]
and $N_z+1$ grid points in the $z-$direction at 
\[ z_j = -z_{\max} + j h_z, \quad j = 0, \ldots N_z, \quad h_z = \frac{2z_{\max}}{N_z}.\]
Gravitational fields are defined at the grid points $(r_i, z_j)$.
We use centred second-order finite differences to discretize the geometry equations in space.
Ghost points are added at the axis and field values at these points set according to their parity in $r$ (\ref{s:appendix.geodetails.bcs}). 
Backward finite differences are used at the outer boundaries $r_{\max} \times [-z_{\max}, z_{\max}]$, $[0, r_{\max}] \times \{-z_{\max}\}$ and $[0, r_{\max}] \times \{z_{\max}\}$.

\subsection{Particle-in-cell method}
\label{s:numerics.pic}
We use a second-order particle-in-cell (PIC) scheme, whereby the Vlasov distribution is represented by a discrete set of particles.
The basic steps of this scheme (explained in detail below) are as follows. 
The gravitational fields are interpolated to the particle locations, which allows the particles to be advanced in time according to the geodesic equations;
simultaneously, the gravitational fields are advanced at the grid points.
Matter source terms are then formed at the advanced time step
by `depositing' the particles onto the grid, and finally, the elliptic equations are solved. 
These steps are iterated in a second-order Runge-Kutta method.

\subsubsection{Initial data}
\label{s:numerics.pic.initial_data}
Before evolving, the initial distribution function must be discretized in phase space. This is done by sampling the distribution by a set of regularly spaced particles. 
In physical space the particles are placed at the centers of the rectilinear grid, while in momentum space $N^v_i$ ($i=1,2,3)$ particles are placed between specified bounds (for examples refer to \Sectionref{s:results.initial_data}). 
Let $h^v_1, h^v_2, h^v_3$ denote the respective distances between particles in each momentum direction. Given an initial distribution function $\hat{f}$\footnote{As explained in \ref{s:appendix.geodetails.iniconstr}, we freely prescribe $\hat{f} := \psi^2 \tf$ for the initial data.} we assign the particles weights $m_\mathrm{p}$ according to 
\begin{equation}
  \label{e:particle_init}
  m_\mathrm{p} = \hat{f}(0, x_\mathrm{p}, v_\mathrm{p})h^v_1 h^v_2 h^v_3,
\end{equation}
where $(x_\mathrm{p}, v_\mathrm{p})$ are the coordinates of the particle locations in 5-dimensional phase space. To reduce computational cost, particles with masses below a problem-dependent threshold are dropped. 
Consistent initial data satisfying the constraint equations are generated as in \ref{s:appendix.geodetails.iniconstr}.  

\subsubsection{Evolution scheme}
\label{s:numerics.pic.evolution_scheme}

The sequence of steps involving the geometry and matter variables in the Runge-Kutta evolution scheme are shown in the pseudo-code below. 
We define the following groups of our fundamental variables:
\bigskip\\
\begin{tabular}{lll}
  $\mbe$: & evolved geometry fields & $s, \tW, \tU, \tX, \tE^A, \tBp$ \\
  $\mbc$: & constrained geometry fields & $\psi, \alpha, \beta^A$ \\
  $\mbm$: & matter variables & $x_\mathrm{p}, v_\mathrm{p}, \tilde{f}_\mathrm{p}$ \\
  $\mbs$: & matter source terms & $\tilde \tau, \tilde J^\phi, \tilde S_A, \tilde \rho_H, \tilde J_A, \tilde S_{AB}$
\end{tabular}
\bigskip\\
Let $\mbfe$ denote the right-hand sides of the evolution equations \eref{e:dts}--\eref{e:dtBp_} for $\mbe$, and $\mbfm$ the right-hand sides of the geodesic equations \eref{e:drdt}--\eref{e:dv3dt} for $\mbm$.
Suppose we are given evolved and constrained geometry variables $\mbe_0, \mbc_0$, matter variables $\mbm_0$ and source terms $\mbs_0$ on time step $t$.
Then the quantities are computed on the next time step $t+\Delta t$ by the following algorithm:
\bigskip\\
\begin{tabular}{lll}
Step & Geometry code & Matter code \\ \hline
  Right-hand side &  $\mbke_1 = \Delta t \, \mbfe(\mbe_0, \mbs_0)$ & $\mbkm_1 = \Delta t \, \mbfm(\mbe_0, \mbc_0, \mbm_0)$ \\
  Evolve & $\mbe_1 = \mbe_0 + \mbke_1$ &  $\mbm_1 = \mbm_0 + \mbkm_1$ \\
  Matter source terms & \mbox{} & $\mbs_1 = \mbs(\mbm_1)$ \\
  Solve constraints & $\mbc_1 = \mbc(\mbe_1, \mbs_1)$ & \mbox{} \\
  \hline
  Right-hand side & $\mbke_2 = \Delta t \, \mbfe(\mbe_1, \mbs_1)$ & $\mbkm_2 = \Delta t \, \mbfm(\mbe_1, \mbc_1, \mbm_1)$ \\
  Evolve & $\mbe_2 = \mbe_0 + \frac 12 (\mbke_1 + \mbke_2)$ &  $\mbm_2 = \mbm_0 + \frac 12 (\mbkm_1 + \mbkm_2)$ \\
  Matter source terms & \mbox{} & $\mbs_2 = \mbs(\mbm_2)$ \\
  Solve constraints & $\mbc_2 = \mbc(\mbe_2, \mbs_2)$ & \mbox{}
\end{tabular}
\bigskip\\
$\mbe_2, \mbc_2, \mbm_2, \mbs_2$ are now a second-order accurate (in $t$) approximation to the solution at time $t+\Delta t$. 
The time step $\Delta t$ is set according to
\begin{equation}
  \Delta t = \lambda \min(h_r, h_z),
\end{equation}  
where we typically choose $\lambda=0.5$ for the Courant-Friedrichs-Lewy number.
Fourth-order Kreiss-Oliger dissipation \cite{Kreiss1973} is added to the right-hand sides of the geometry evolution equations.

\subsubsection{Interpolation}
\label{s:numerics.pic.interpolation}
The steps 
\emph{Right-hand side}
and \emph{Matter source terms} involve interpolating between the particles and the grid.  Interpolation is performed using a linear interpolation kernel. 
In the \emph{Right-hand side}
step (computation of $\mbkm$) each field $u_\mathrm{p}$ at the particle location $(r_\mathrm{p}, z_\mathrm{p})$ is computed from the grid quantity $u$ by  
\begin{equation}
  u_\mathrm{p} = \sum_{(i,j)} W\left(\frac{|r_i - r_\mathrm{p}|}{\Delta r}\right)W\left(\frac{|z_j - z_\mathrm{p}|}{\Delta z}\right)u(r_i, z_j),
\end{equation}
where the interpolation kernel is given by 
\[
  W(x) = \cases{1 - x, & $0 \le |x| < 1$, \\ 
  0 & otherwise.}
\]
In practice the sum over grid indices $\sum_{(i,j)}$ is shortened to a sum over indices with coordinates satisfying $|x^A_i - x^A_\mathrm{p}|/{\Delta x^A} \le 1$. 
The same interpolation (with $u$ and $u_\mathrm{p}$ reversed) is done when constructing the matter source terms. 

\subsubsection{Constraint solver}
\label{s:numerics.pic.constraint_solver}
Linear elliptic equations give rise to sparse linear systems of equations when discretized using finite differences, which we solve using a direct sparse matrix solver (\textsc{Umfpack}). 
The only nonlinear equation in our system is the Hamiltonian constraint \eref{e:hamcons} (or \eref{e:hamconsini} for the initial data) on $\psi$, for which we use an outer Newton-Raphson iteration, at each step solving the linear equations as above.

\subsubsection{Boundary treatment for particles}
\label{s:numerics.pic.boundary_treatment}
Care must be taken in the above steps to handle particles which have left the domain. 
Particles that leave the domain at the outer boundaries are simply dropped. 
In axisymmetry, the gravitational field within a cylinder of finite radius is influenced by energy-momentum outside of this radius. 
Therefore, dropping particles as just described is a source of numerical error. 
In our numerical experiments the domain is chosen large enough to minimize the loss of particles while maintaining reasonable computing times.
In cases, our simulations have been validated by comparing with runs on larger domains. 

On the other hand, some particles may achieve $r$-values which are less than zero. To deal with such particles we implement a particle reflection for particles with $r$-values less than $h_r/4$, which is one half of the smallest grid index. This transformation reflects offending particles about $h_r/4$, and reverses the radial momentum of ingoing particles, 
\[ 
  r \to \frac{h_r}{4} + \left|r- \frac{h_r}{4}\right|, \qquad v_r \to - v_r, \qquad \mbox{ for particles with } v_r < 0. 
\]
Due to troublesome factors of $r^{-1}$ in some of the matter source terms \eref{e:tautildeframe}--\eref{e:Sddtildeframe}, we typically choose initial data in such a way that no particles have zero angular momentum. 
This helps limit the number of particles which approach or seek to ``cross through'' the axis.


\subsection{Horizon finder}  \label{s:numerics.horizon_finder}

We detect black hole formation by looking for an apparent horizon.
This is defined as the outermost 2-surface in a spatial slice whose outgoing null expansion vanishes.
In our dimensional reduction, such a surface corresponds to a curve in the $(r,z)$ plane.
We assume that this curve can be parametrized by the spherical polar angle $\theta$ as
\begin{equation}
  r = R(\theta) \sin\theta, \quad z = R(\theta) \cos\theta,
\end{equation}
where $R$ is the spherical polar radius.
(One should remark that this is a certain restriction as not all apparent horizons can be parametrized in this way.)
The function $R(\theta)$ obeys a second-order ODE derived in \ref{s:appendix.diagnostics.horizons}.
Smoothness of the apparent horizon on the axis implies the boundary conditions $R'(0) = R'(\pi) = 0$.
We solve this two-point boundary value problem using the shooting method.
Even if the shooting method fails so that an apparent horizon does not exist, we compute the minimum value of $R'(\pi)$ for a range of initial values $R(0)$.
We have found this quantity very useful in monitoring the approach to apparent horizon formation (figure \ref{f.bh_formation}) \cite{Garfinkle2001,Rinne2008}.

If an apparent horizon is found (so that $R'(\pi) = 0$), we compute its mass and angular momentum as detailed in \ref{s:appendix.diagnostics.horizons}.
We normally stop a numerical simulation when an apparent horizon forms as we experience increasingly large violations of the constraints and of mass and anguar momentum conservation starting just before this time.


\subsection{Convergence tests}
\label{s.numerics.convergence}
As discussed above the momentum constraint equations are not solved in our evolution scheme, and the residuals of these equations can be monitored for code accuracy. We present convergence results for these quantities below. Let $\mathcal C_i$ ($i = r,z,\phi$) denote the right hand-sides of equations \eref{e:momconsr}, \eref{e:momconsz}, and \eref{e:gercons}
In comparing the residual across evolutions at different resolutions we compute the normalized quantity
\begin{equation}
  \label{e.normalized_residuals}
  \overline{\mathcal C}_i := \frac{\|\mathcal C_i\|_{L^2}}{(1 + \sum_{k=1}^{K_i}\|\mathcal C_i^k\|_{L^2})},
\end{equation}
where $\mathcal C_i^k$ is the $k^\mathrm{th}$ term in the $i^\mathrm{th}$ constraint equation, and $K_i \in \mathbb N$ is the number of terms. 
The norm $\|\cdot\|_{L^2}$ denotes a discrete 2-norm.
The quantity $\overline{\mathcal C}_i$ is thus bounded between $0$ and $1$.

We present convergence results for two of the families, one rotating and one nonrotating, 
discussed below in \Sectionref{s:results}. For each, the key parameters are taken as follows. Low resolution: $N_r = 250, \, N_z = 500, \, N_v = 12$. Mid resolution: $N_r = 354, \, N_z = 708, \, N_v = 18$. High resolution: $N_r = 500, \, N_z = 1000, \, N_v = 24$. 
In each case, $N^v_1 = N^v_2 = N^v_3 =: N_v$.
The grid resolutions increase approximately by a factor of $\sqrt{2}$, while the number of particles in each momentum direction is chosen by multiplying by the same factor and taking the nearest larger even integer. In addition, within each family the amplitude is adjusted until the initial Dain mass (section \ref{s:appendix.diagnostics.conserved}) is within a tenth of a percent. Total particle numbers in each case are roughly $1\times 10^6, 9\times 10^6$, and $50\times10^6$ respectively.

\begin{figure}[htb]
  \centering
  \begin{subfigure}[b]{0.5\linewidth}
    \includegraphics[width=\textwidth]{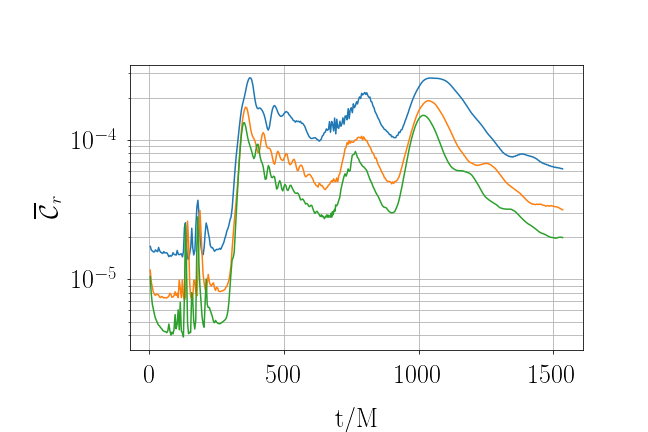}
  \end{subfigure}\qquad
  \begin{subfigure}[b]{0.5\linewidth}
    \includegraphics[width=\textwidth]{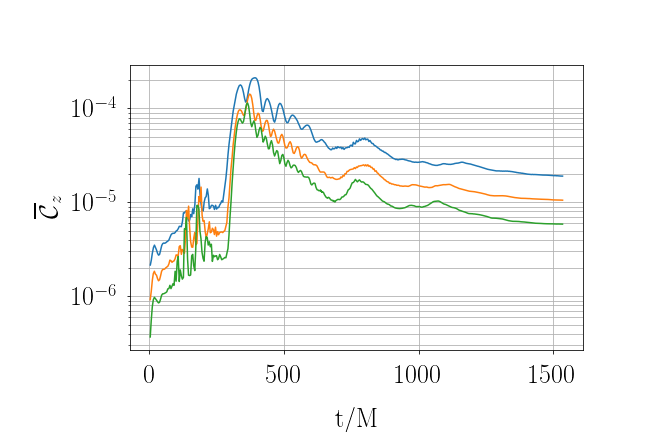}
  \end{subfigure}
  \caption{\label{f.constraint_convergence_nrf3} Normalized residuals for the momentum constraint quantities for data family NR3. 
  In the left panel $\overline{\mathcal C}_r$, and in right panel $\overline{\mathcal C}_z$.
  Low (blue), mid (orange), and high (green) resolution, as defined in the text, are shown from top to bottom.
  }
\end{figure}

For the non-rotating data (taken to from the NR3 family in table \ref{t.initial_data}) the initial distribution undergoes a slight contraction before $t = 50M$, after which it expands and eventually disperses. 
The residuals (see figure \ref{f.constraint_convergence_nrf3}) can be seen to increase during the contraction of the matter and remain constant, or even decrease as it expands. 
Noise in the residuals is largely due to the matter terms in the constraint equations.

Similar results are seen in the rotating case, in which data belongs to the R7 family (see table \ref{t.initial_data} below). 
Figure \ref{f.constraint_convergence_rf7} shows the normalized constraint quantities $\overline{\mathcal C}_r$, $\overline{\mathcal C}_z$, and $\overline{\mathcal C}_\phi$ for the same three resolutions as with the non-rotating case. 

\begin{figure}[htb]
  \centering
  \begin{subfigure}[b]{0.5\linewidth}
    \includegraphics[width=\textwidth]{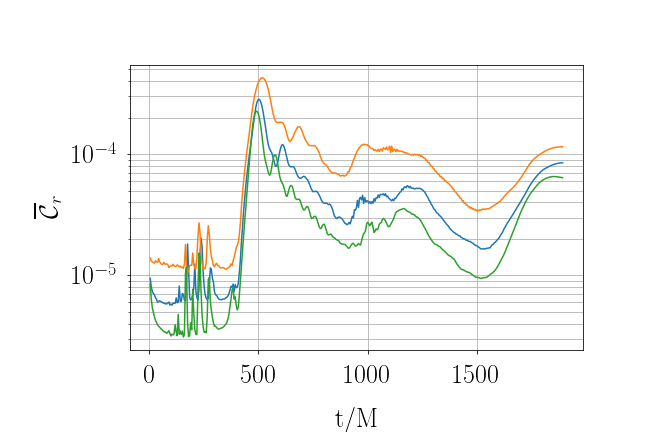}
  \end{subfigure}\qquad
  \begin{subfigure}[b]{0.5\linewidth}
    \includegraphics[width=\textwidth]{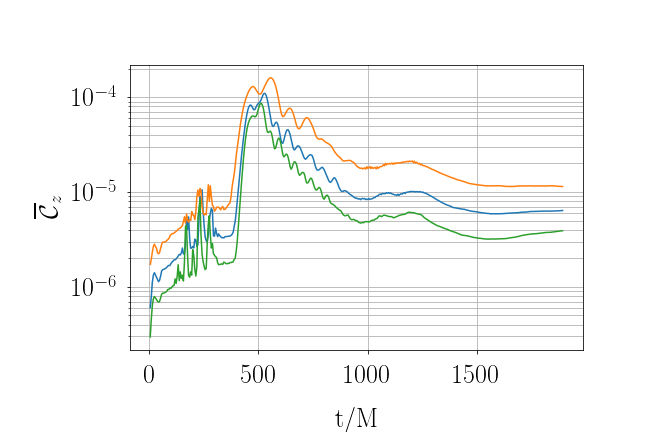}
  \end{subfigure}\\
    \begin{subfigure}[b]{0.5\linewidth}
    \includegraphics[width=\textwidth]{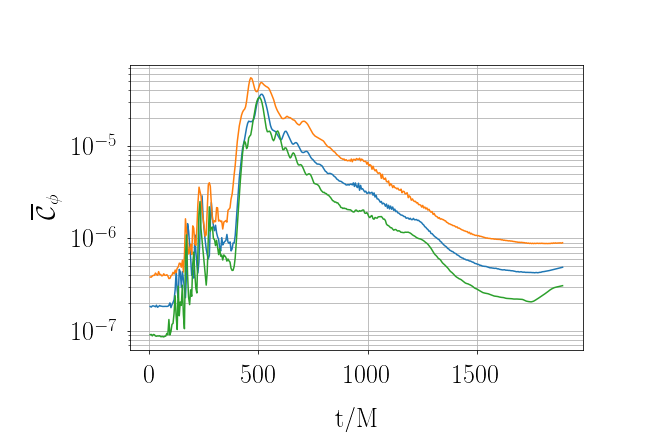}
  \end{subfigure}
    \caption{   
    \label{f.constraint_convergence_rf7}
    Normalized residuals for the momentum constraint quantities for data family R7.
    Low (blue), mid (orange), and high (green) resolution, as defined in the text, are shown from top to bottom.
    }
\end{figure}

For a strictly grid-based second-order finite-difference scheme, one
would expect the residuals for the different resolutions (which are
spaced by factors of $\sqrt{2}$) to be spaced by factors of $2$. 
In the vacuum case, where no particles are present, our code produces such an expected result.
Convergence of particle-in-cell codes in the context of the Einstein equations has been investigated by Rein and Rodewis \cite{ReinRodewis} in the spherically symmetric setting.
The authors find that using linear interpolation there is a correction to perfect quadratic convergence depending on the ratio of the matter cell size to the grid spacing. 
While these results suggest that quadratic convergence is unlikely in our setting, our numerical results indicate that we do see convergence which is close to $2$ and that our simulations, which run with a resolution approximately that of `Mid resolution', are in the convergent regime.
The constraint violation quantities given by \eref{e.normalized_residuals} for the runs presented in section \ref{s:results} remain less than $0.01$ throughout the evolution. 


\section{Numerical results}
\label{s:results}


\subsection{Initial data}
\label{s:results.initial_data} 

The initial data for the rescaled density function 
$\hat f= \psi^2 \tilde f = \psi^8 e^{2rs} f$ (cf. \ref{s:appendix.geodetails.iniconstr}) used in the simulations below is given by 
\begin{eqnarray}
\fl\mathring{f}(r,z,v_1,v_2,v_3)=A\big((r^{\max}-r)_+(r-r^{\min})_+(z^{\max}-z)_+(z-z^{\min})_+\big)^2\nonumber\\
\big((v_1^{\max}-v_1)_+(v_1-v_1^{\min})_+(v_2^{\max}-v_2)_+(v_2-v_2^{\min})_+\big)^2\nonumber\\
\big((v_3^{\max}-v_3)_+(v_3-v_3^{\mathrm{gapmax}})_+ + (v_3-v_3^{\min})_+(v_3^{\mathrm{gapmin}}-v_3)_+\big)^2.
\label{e:initialdata}
\end{eqnarray}
Here $A, r^{\min}, r^{\max}, z^{\min}, z^{\max}, v_1^{\min}, v_1^{\max}, v_2^{\min}, v_2^{\max}, v_3^{\min}, v_3^{\max}, v_3^{\mathrm{gapmin}}$ and $v_3^{\mathrm{gapmax}}$ 
are parameters and $(x)_+=x$ if $x\geq 0$ and $(x)_+=0$ if $x<0$. We call the parameter $A$ the amplitude. It is nonnegative 
which implies that $\mathring{f}\geq 0$. 
Furthermore, we always choose $z^{\min}=-z^{\max}<0$. 
In the spatial coordinates $r, z$ and $\phi$, the configuration described by $f$ is topologically a torus when $r^{\min}>0$, which we always take.
The reason is that numerically the axis gives rise to difficulties for the matter quantities in the coordinates we are using. 
Hence, as long as there is no matter close to the axis we have better control 
of the accuracy of the simulation. 
On the other hand we cannot choose $r^{\min}$ too large since our numerical domain is finite and matter should not leave the domain too quickly. 
In this context, let us compare with the spherically symmetric case.
In that setting the solution outside the support of the matter is the Schwarzschild solution and it is sufficient that the numerical grid covers the support of the matter. 
In axisymmetry the solution outside the support is not explicitly known and we need a rather large domain, even if the matter (initially) only occupies a small part of the domain. 
The role of the parameters $v_3^{\mathrm{gapmin}}$ and $v_3^{\mathrm{gapmax}}$ is to make sure that the angular momentum of the particles are strictly bounded away from zero. 
The reason is again to try to avoid that particles come close to the axis. If gravity is strong enough this is not possible to avoid, but in a weak gravitational field non-zero angular momentum for the particles will keep them away from the axis. 
If $v_3^{\mathrm{gapmin}}=-v_3^{\mathrm{gapmax}}$ and $v_3^{\mathrm{min}} = -v_3^{\mathrm{max}}$
then the total angular momentum is zero since there are equally many particles rotating in the two directions and we obtain a non-rotating configuration. 
If the parameters do not satisfy this condition the total angular momentum is non-zero and we have a rotating configuration. 

In the sections below we present examples of solutions launched by such data with the following parameters.  

\begin{table}[h!]
\begin{center}
  \begin{tabular}{ |c|c|c|c|c|c|} 
    \hline
    Family & $r$ & $z$ & $v_1$ & $v_2$ & $v_3$ \\   
    \hhline{|=|=|=|=|=|=|} 
    NR3 
    & $[8,14]$ & $[-3,3]$ 
    & $[-0.2, 0]$ & $[-0.1, 0.1]$ 
    & $[-0.2, -0.1] \cup [0.1, 0.2]$ \\     
    \hline
    R7 
    & $[8,14]$ & $[-3,3]$ 
    & $[-0.2, 0]$ & $[-0.1, 0.1]$ 
    & $[-0.19, -0.1] \cup [0.1, 0.2]$ \\    
    \hhline{|=|=|=|=|=|=|} 
    NR4 
    & $[8,14]$ & $[-3,3]$ 
    & $[-0.32,-0.12]$ & $[-0.1, 0.1]$
    & $[-0.2,-0.1] \cup [0.1, 0.2]$ \\  
    \hline
    NR2 
    & $[4,12]$ & $[-3,3]$ 
    & $[-0.5, 0]$ & $[-0.5, 0.5]$ 
    & $[-0.5, 0.5]$ \\  
    \hline
    R4 
    & $[4,12]$ & $[-3,3]$ 
    & $[-0.5, 0]$ & $[-0.5, 0.5]$ 
    & $[-0.45, 0.55]$ \\
    \hline
  \end{tabular}
  \caption{\label{t.initial_data}
    Parameters from \ref{e:initialdata} indicating support of distribution in phase space.
    The first two families have critical binding energy $E_b^* > 0$, the last three $E_b^* < 0$.
  }
  \end{center}
\end{table}
We have essentially two sets of families with rather different spatial extents but in each of these families the parameters $r^{\min}, r^{\max}$ and $z^{\max}$ are fixed. 
By varying the amplitude and the parameters in the momentum space we generate initial data which have different characteristics. 
The characteristics we have in mind are the physical quantities that we have found to be the most important ones for determining the asymptotic behaviour of a solution. 
These are the ADM mass $M$, the ratio $|J|/M^2$ and the fractional binding energy $E_b$, which is defined by
\begin{equation}\label{binding-energy}
E_b=\frac{M_0-M}{M_0},
\end{equation}
where $M_0$ is the rest mass. See \ref{s:appendix.diagnostics.conserved} for the definition of the quantities $M, M_0$ and $J$. 

Families NR2, NR3 and NR4 have zero net angular momentum, while R4 and R7 both have non-zero angular momentum. 
The data families are also distinguished by the sign of the binding energy for the critical data $E_b^*$. 
For NR3 and R7 we have $E_b^* > 0$, whereas for NR2, NR4 and R4 we have $E_b^* < 0$.

\subsection{End states}

One aspect which we have focused on in the present 
work is to investigate which types of asymptotic behaviour in the evolution are possible, and we have tried to understand which 
quantities are important for distinguishing these types. 
We have characterized the general asymptotic behaviour into the following three types:
\begin{enumerate}
  \item Complete dispersion
  \item Oscillatory behaviour
  \item Black hole formation
\end{enumerate}
By complete dispersion we mean that there is a monotonically increasing function $R=R(t)$, where $R(t)\to \infty$ as $t\to\infty$, 
and a time $t_\mathrm{d}$, such that $f(t,r,z,v_k)=0$ for $(r,z)$ satisfying $\sqrt{r^2+z^2}\geq R(t)$ when $t>t_\mathrm{d}$. 
Hence, all matter eventually leaves a ball of radius $R(t)$ for any $t$. 
There are cases where the matter splits into different parts during the evolution and where some part disperses. 
These cases we characterize as belonging to one of the two other types, namely the oscillatory behaviour or black hole formation.  
By oscillatory behavior we include the possibilities that the end state is periodic, asymptotically periodic in a sense (cf. \cite{Makino2016}), or is a stationary state.
The oscillations we observe are typically damped.
Likewise, if some part collapses to a black hole whereas some part disperses, we characterize the asymptotic behaviour to belong to the third type; formation of a black hole. 
When a black hole forms it is, in cases, preceded by damped oscillatory behaviour (see figures \ref{f.rpeak_rpmin.rf7} and \ref{f.rpmin.Eb_positive} below).
A possible fourth type would be that some matter 
collapses whereas some matter enters the oscillatory phase but in our simulations we have not observed such a behaviour. 
(One should note however that we stop the simulation as soon as an apparent horizon forms.)

\begin{figure}[h]
  \centering
  \includegraphics[width=0.5\textwidth]{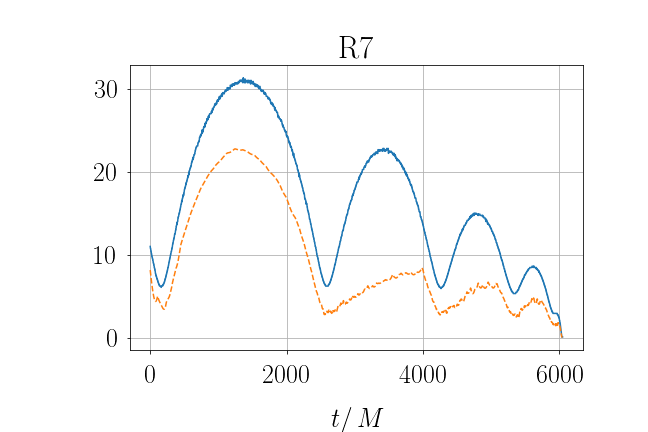}
  \caption{ \label{f.rpeak_rpmin.rf7}
    Radius of peak density (blue, solid), and minimum value of the radial coordinate $r_{p}$ (orange, dashed) for R7 supercritical data with amplitude $A=9.6 \times 10^8$.
  }
\end{figure}
A crucial quantity for determining the asymptotic behaviour of the solutions is the binding energy $E_b$. In Proposition 1 below we show that complete dispersion 
implies that $E_b\leq 0$. 
In spherical symmetry a slightly stronger result holds, namely it is shown in \cite{Andreasson2006} that dispersion 
implies that $E_b<0$. Whether or not dispersion implies that $E_b$ is strictly negative also in axisymmetry is an open question. 
In the case of positive binding energy we find numerically that it leads to oscillatory behaviour (in the general sense defined above) or black hole formation, cf. the discussion in section \ref{s.results.Eb_positive}.

A non-positive binding energy is clearly a necessary condition for complete dispersion in view of Proposition 1. It is however not a sufficient condition. 
In fact several end states are possible. In spherical symmetry there are rigorous results which show that at least two different end-states may occur 
in that case. 
Firstly, in \cite{Rein1992} it is shown 
that sufficiently small initial data leads to dispersion. An example of initial data that satisfy 
the conditions in \cite{Rein1992} is given by $\mathring{f}=Ag$ where $g$ is a given spherically symmetric density function with compact support and where $A>0$ is 
a constant. From the expression of the binding energy in spherical symmetry, cf. \cite{Andreasson2006}, it is clear that by taking $A$ sufficiently small the binding energy is negative. 
Clearly, there exist initial data with negative binding energy which leads to dispersion. Similarly, by using arguments in \cite{Andreasson2012} 
one can prove that there exist initial data with $E_b<0$ such that black holes form in the evolution. 
In the present simulations we have launched small initial data 
with negative binding energy, which leads to dispersion, and in the study of critical collapse below we have constructed data with negative binding energy 
which lead to black hole formation, cf. section \ref{s.results.Eb_negative}.
We have also encountered initial data with negative binding energy where parts of the matter are ejected in a violent explosion, and where the remaining part 
has positive binding energy. This part then enters a phase of oscillatory behaviour. 
This case is also discussed in section \ref{s.results.Eb_negative}.

Another quantity which is crucial for the asymptotic behaviour is the ratio $|J|/M^2$. The cosmic censorship conjecture implies that a black hole cannot form with $|J| > M^2$. By launching data tailored for gravitational collapse we find very interesting dynamics: an initial implosion 
is followed by a series of oscillations, where particles are ejected, and finally a black hole forms. We have numerical indications that the remaining matter 
satisfies $|J| < M^2$ when the black hole forms. 
See section \ref{s.results.super_rotating} for discussion. 

Let us finish this subsection by the following proposition. 
\begin{proposition}
  \label{prop.complete_dispersion}
  If complete dispersion occurs then the binding energy $E_b\leq 0$. 
\end{proposition}
  
\textbf{Proof: }We assume that a regular global solution exists to the Einstein-Vlasov system which satisfies the specified fall off conditions. 
We assume that the fractional binding energy is strictly positive and we show that this leads to a contradiction. 
Accordingly we set 
\[
0<\epsilon:=M_0-M.
\]
Next we note that the ADM mass can be split into two non-negative terms where the first term only contains field quantities and the second term contains the energy density $\rho_H$. We get 
\begin{eqnarray}
  M_0-M
  &\leq M_0-\int_{-\infty}^\infty \rmd z \int_0^\infty \fourth \kappa \psi^{-2} \tilde \rho_H r \, \rmd r \nonumber\\ 
  &= M_0
    -\int_{-\infty}^\infty \rmd z \int_0^\infty 2\pi \tilde{\rho}_H r \, \rmd r
    +\int_{-\infty}^\infty \rmd z \int_0^\infty2\pi (1-\tilde{\psi}^{-2})\rho_H r \, \rmd r\nonumber\\
  &= 2\pi \int_{\Sigma_{t}}  \left(  \int_{\mathbb R^3}\big(1-\sqrt{1+v_1^2+v_2^2+v_3^2}\big) \tilde{f}(r, z, v) \, \rmd^3 v \right) r \, \rmd r \, \rmd z\nonumber\\
  & \quad  +\int_{-\infty}^\infty \rmd z \int_0^\infty2\pi (1-\tilde{\psi}^{-2})\rho_H r \, \rmd r\nonumber\\
  &\leq \int_{-\infty}^\infty \rmd z \int_0^\infty2\pi (1-\tilde{\psi}^{-2})\rho_H r \, \rmd r. \label{binding}
\end{eqnarray}
Next we use that $f$ and consequently $\rho_H$ have support in the domain $r^2+z^2\geq R^2(t)$ for $t>t_\mathrm{d}$. For large $t$ the last term can be estimated by
\begin{equation}\label{asymp}
\left|\int_{-\infty}^\infty \rmd z \int_0^\infty2\pi (1-\tilde{\psi}^{-2})\rho_H r \, \rmd r\right|\leq \frac{CM}{R(t)},
\end{equation}
where we used that $|1-\psi^{-2}|\leq\frac{C}{R(t)}$ when $r^2+z^2\geq R^2(t)$. 
For sufficiently large $t$ we have 
\[
\frac{CM}{R^2(t)}<\frac{\epsilon}{2}, 
\]
and from \ref{binding} we thus get 
\begin{equation}
M_0-M\leq \frac{\epsilon}{2}. 
\end{equation}
This leads to the desired contradiction and the proof is complete. 
\begin{flushright}
$\Box$
\end{flushright}


\subsubsection{Formation of black holes}

\begin{figure}[h]
  \centering
  \begin{subfigure}[b]{0.49\linewidth}
    \centering
    \includegraphics[width=\textwidth]{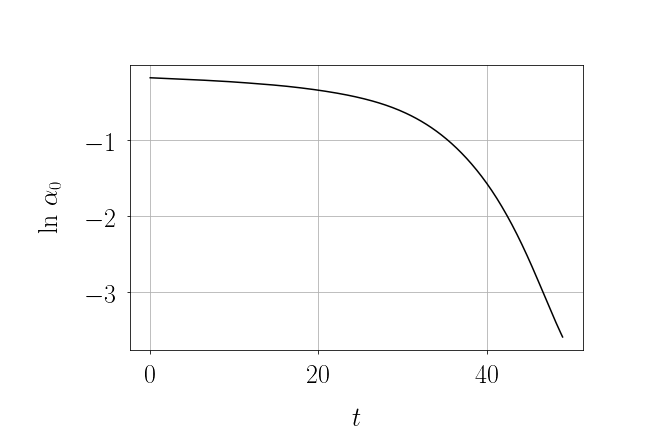}
  \end{subfigure} 
  \begin{subfigure}[b]{0.49\linewidth}
    \centering
    \includegraphics[width=\textwidth]{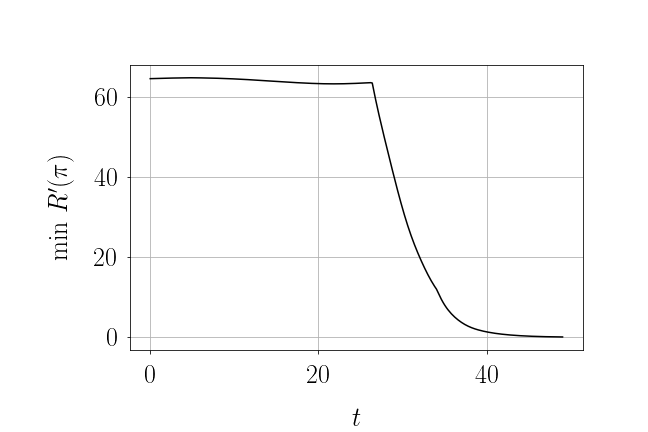}
  \end{subfigure}\\
  \begin{subfigure}[b]{0.49\linewidth}
    \centering
    \includegraphics[width=\textwidth]{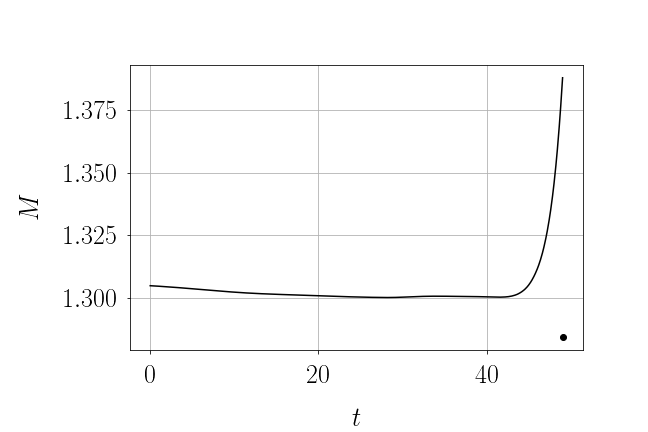}
  \end{subfigure} 
  \begin{subfigure}[b]{0.49\linewidth}
    \centering
    \includegraphics[width=\textwidth]{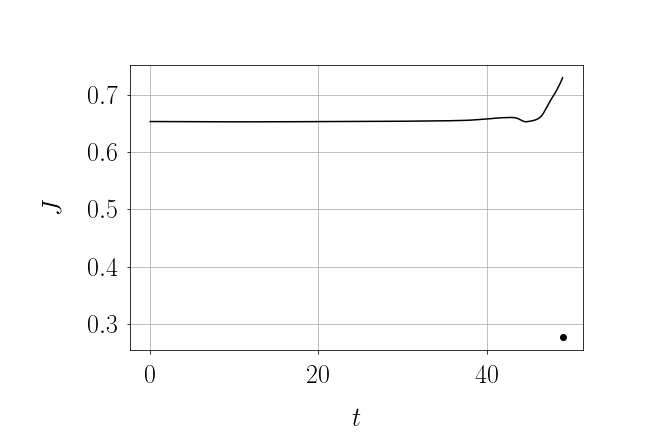}
  \end{subfigure} 
  \caption{ \label{f.bh_formation}
    Black hole formation in a strongly supercritical evolution of family R4 (amplitude $A=0.1$).
    See the text for a description of the quantities shown.
    }
\end{figure}

For large amplitudes in our initial data families, black holes form generically. 
In figure \ref{f.bh_formation} this is illustrated for the 
rotating family R4.
Typical indications of black hole formation are the collapse of the lapse in the  maximal slicing we use (shown is $\ln\alpha_0$, the logarithm of the lapse at the origin) and a rapid decrease of the quantity $\min R'(\pi)$ in our horizon finder (section \ref{s:numerics.horizon_finder}), which becomes zero when the apparent horizon forms. 
Also shown in figure \ref{f.bh_formation} are the ADM (or Dain) mass $M$ \eref{e:DainMass} and the angular momentum $J$ \eref{e:J}, as computed on our finite computational domain.
These are reasonably well conserved during the evolution but tend to increase just before the horizon forms.
The numerical accuracy suffers in this regime as a result of the steep gradients that develop due to the slice stretching.

The mass $M_H$ \eref{e:M_H} and angular momentum $J_H$ \eref{e:J_H} of the apparent horizon, when it forms, are shown as dots in figure \ref{f.bh_formation}. 
In accordance with the Penrose inequality, the horizon mass is smaller than the ADM mass.
However, due to the numerical violations of mass and angular momentum conservation just before the black hole forms, we cannot determine the black hole mass and angular momentum accurately in the current version of our code.

\subsection{Critical behaviour}
We investigate critical collapse for the following families of initial data:  
\begin{enumerate}
  \item Non-rotating data where the critical solution has positive binding energy 
  \item Non-rotating data where the critical solution has negative binding energy
  \item Rotating data satisfying $|J| < M^2$ with positive binding energy
  \item Rotating data satisfying $|J| < M^2$ with negative binding energy
\end{enumerate}
We note a difference in the dynamics of supercritical solutions in cases where the critical solution has positive versus negative binding energy. 
The positive binding energy case is presented in section \ref{s.results.Eb_positive} below, and in section \ref{s.results.Eb_negative} we present the negative binding energy case.
We also investigate initial data for which $|J| > M^2$.
Our findings are briefly discussed in section \ref{s.results.super_rotating}.

We observe that the time $t_\mathrm{H}$ when the apparent horizon forms in supercritical evolutions obeys approximately the scaling law
\begin{equation} \label{e:tscaling}
  \frac{t_\mathrm{H}}{M_*} = -\sigma \ln|A - A_*| + \const.
\end{equation}
We determine the critical amplitude $A_*$ and the scaling exponent $\sigma$ by performing a nonlinear fit to our data for several amplitudes $A$ and take $M_*$ to be the ADM mass of the data with amplitude closest to $A_*$.

This scaling law can be understood if we assume that the critical solution is an unstable attractor with a single unstable mode in linear perturbation theory. 
A simple dimensional analysis \cite{Gundlach2007} shows that the time the solution spends in the vicinity of the critical solution obeys a scaling law of the form \eref{e:tscaling}, where $\sigma=1/\lambda_0$ is given by the (positive) real part $\lambda_0$ of the eigenvalue corresponding to the unstable mode.
From our simulations it appears that the approach to the critical solution and the departure from it until the black hole forms are very similar for all near-critical evolutions.  
Hence, using the time $t_\mathrm{H}$ (which is much easier to determine) instead of the lifetime of the critical solution in \eref{e:tscaling}, will only add a constant offset.

The results presented here were run on domains of size $[0, 40]\times[-40, 40]$ with $N_r = 300, N_z = 600$, or $[0, 60]\times[-30, 30]$ with $N_r = 400, N_z = 400$, and with $16$ particles in each momentum direction. 
The total number of particles, which depends generally on the support of the matter, is between $3$ and $4$ million. 


\subsubsection{Families where the critical solution has 
positive binding energy
}
\label{s.results.Eb_positive}

In this section we report results from families NR3 and R7 from table \ref{t.initial_data}.
The family R7 is highly rotating in the near critical regime, $|J|/M^2 \approx 0.9$, while NR3 is nonrotating. 
We have also investigated data with more moderate rotation. 
Despite the large difference in the angular momentum ratio in these data families we see similar dynamics in the near-critical behaviour.

\begin{figure}[h]
  \centering
  \begin{subfigure}[b]{0.49\linewidth}
    \centering
    \includegraphics[width=\textwidth]{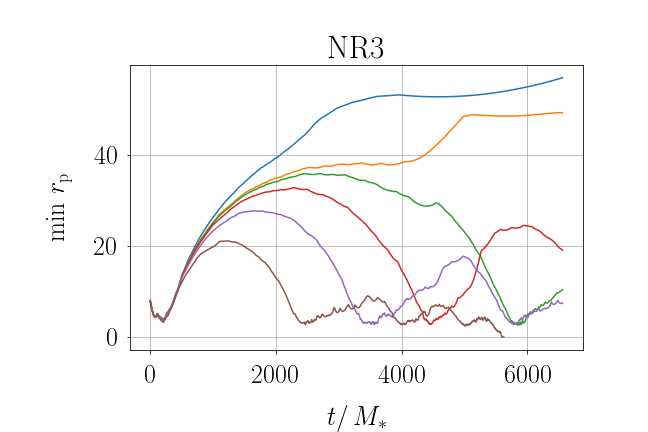}
  \end{subfigure}
  \begin{subfigure}[b]{0.49\linewidth}
    \centering
    \includegraphics[width=\textwidth]{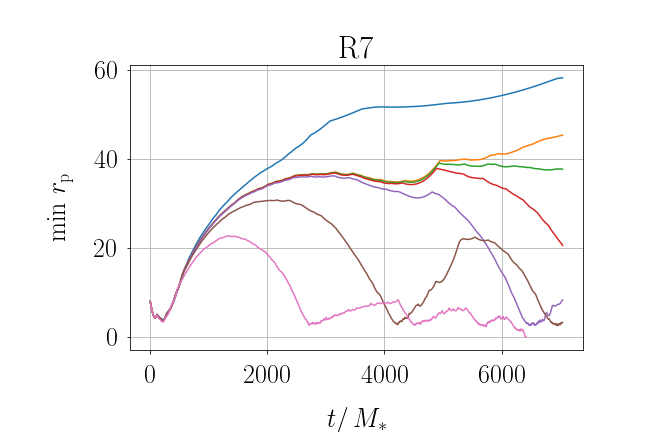}
  \end{subfigure}
  \caption{ \label{f.rpmin.Eb_positive}
    Spatial location of the matter for a few near-critical evolutions of family NR3 (amplitudes $A = (7.0, 7.19, 7.22, 7.3, 7.5, 8.0) \times 10^{8}$ from the top to the bottom curve), and 
    family R7 (amplitudes $A= (8.6, 8.807, 8.808, 8.81, 8.82, 9.0, 9.6)\times 10^{8}$ from the top to the bottom curve).
    Shown is the minimum value of the radial coordinate $r_{p}$ of the particles.
  }
\end{figure}

In all evolutions there is a short initial collapse phase followed by an expansion lasting a time that depends on the data. 
Subsequently, solutions with $A > A_*$ undergo a series of damped oscillations, which terminate in the final collapse to a black hole.
These damped oscillations can be seen in the radial coordinate of the peak density as in figure \ref{f.rpeak_rpmin.rf7}, in  
\href{https://youtu.be/ULIsHGKMIm8}{movie 1}
\footnote{Links to the individual movies at \url{https://youtube.com/channel/UCtxVV46V-kViqNH6hw_JE2w} are included in the electronic version of the article.}
of the energy density $\tilde \rho_H$ and in the minimum radial coordinate of the particles as shown in figure \ref{f.rpmin.Eb_positive}.
As the critical point is approached from above, the amplitude and period of the first oscillation increase.
Both amplitude and period of subsequent oscillations decrease with time.

\begin{figure}[h]
  \centering
  \begin{subfigure}[b]{0.5\linewidth}
    \centering
    \includegraphics[width=\textwidth]{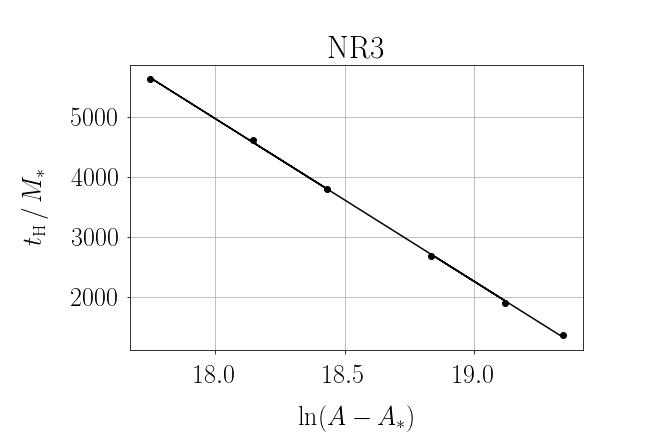}
    \label{f:tscaling_nrf3}
  \end{subfigure}\qquad
  \begin{subfigure}[b]{0.5\linewidth}
    \centering
    \includegraphics[width=\textwidth]{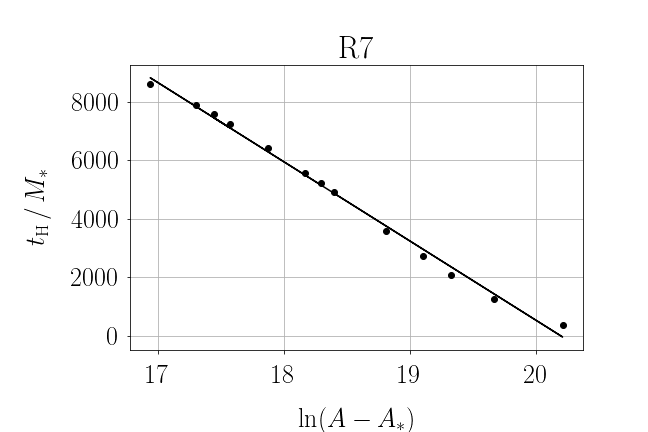}
    \label{f:tscaling_rf7}
  \end{subfigure}
  \caption{ \label{f.tscaling.Eb_positive}
    Lifetime scaling for initial data families NR3 and R7 with positive binding energy. 
    Shown is the time $t_\mathrm{H}$ when the apparent horizon forms (rescaled by the ADM mass $M_*$ of near-critical data) for several amplitudes $A$, as well as the best fit of the form \eref{e:tscaling}.
    The scaling exponents are found in table \ref{t.near_critical}.
  }
\end{figure}

After a finite number of such damped oscillations the supercritical solutions collapse to a black hole. 
The time to horizon formation is seen to scale with $|A - A_*|$.
Figure \ref{f.tscaling.Eb_positive} shows excellent agreement with the scaling law \eref{e:tscaling} for both NR3 and R7.
The critical amplitudes, scaling exponents, as well as the mass, angular momentum and binding energy of the near critical state are shown in table \ref{t.near_critical}.

\begin{table}[h!]
  \begin{center}
    \begin{tabular}{|c|c|c|c|c|c|c|} 
      \hline
      Family & $A_*$ & $\sigma$ & $M_*$ & $J_*$ & $E_b^*$ \\ 
      \hhline{|=|=|=|=|=|=|=|} 
      NR3 & $(7.49 \pm 0.05) \times 10^8$ & $(2.70 \pm 0.14) \times 10^3$ 
          & $0.442$ & 0 & $1.1 \times 10^{-1}$ \\   
      \hline
      R7  & $(9.02 \pm 0.04) \times 10^8$ & $(2.71 \pm 0.15) \times 10^3$ 
          & $0.426$  
          & $0.166$ 
          & $9.8 \times 10^{-3}$ \\ 
      \hhline{|=|=|=|=|=|=|=|} 
      NR4 & $(9.7862 \pm 0.0005) \times 10^8$ & $(8.08 \pm 0.10) \times 10^2$
          & $0.576$ & 0 & $-3.8 \times 10^{-2}$ \\   
      \hline
      NR2 & $(7.315 \pm 0.002) \times 10^{-2}$ & $(5.1 \pm 0.7) \times 10^2$ 
          & $1.01$ & 0 & $-1.1 \times 10^{-3}$ \\   
      \hline
      R4 & $(7.494 \pm 0.001) \times 10^{-2}$ & $(2.5 \pm 0.6) \times 10^2$ 
          & $1.03$ & 0.49 & $-1.1 \times 10^{-3}$ \\   
      \hline
    \end{tabular} 
    \caption{ \label{t.near_critical}
      Near-critical characteristics of initial data as determined by lifetime scaling of supercritical data.
    } 
    \end{center}
  \end{table}

The lifetime scaling for supercritical data indicates that the critical solution has positive binding energy. 
Decreasing the amplitude further we find data with negative binding energy. 
All such negative binding energy data are found to disperse in the sense that after a brief contraction phase, all of the matter becomes outward moving and completely leaves the computational domain, and the fields decay to flat space. 
There are also subcritical solutions with positive binding energy. 
In light of proposition \ref{prop.complete_dispersion}, these solutions cannot disperse. 
Yet, the subcritical and positive binding energy solutions we have evolved are seen to leave the computational domain.
We believe that this is because the domain is too small; this issue is further discussed below in section \ref{s:conclusion}.

\begin{figure}[h]
  \centering
  \begin{subfigure}[b]{0.49\linewidth}
    \centering
    \includegraphics[width=\textwidth]{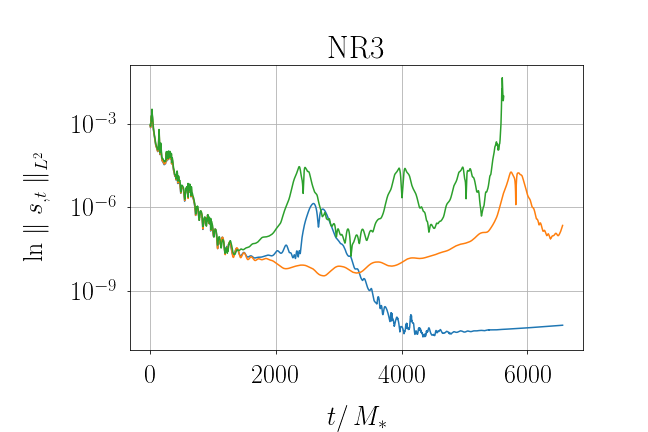}
  \end{subfigure}
  \begin{subfigure}[b]{0.49\linewidth}
    \centering
    \includegraphics[width=\textwidth]{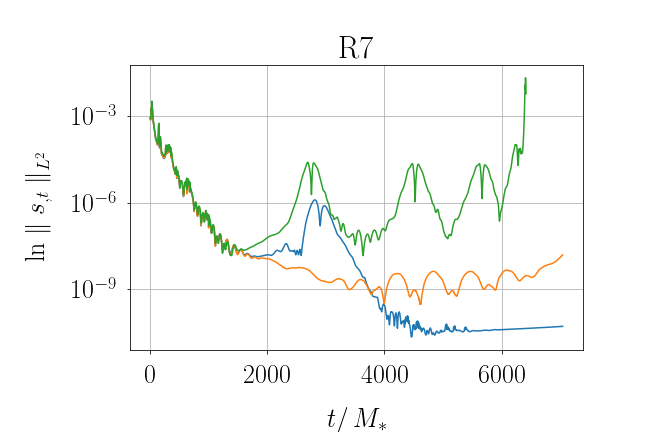}
  \end{subfigure}
  \caption{ \label{f.dts.Eb_positive}
    $L^2$-norm of the log of the right-hand side of the $s$-evolution equation \eref{e:dts} as a function time rescaled by the mass of the critical solution; 
    shown for several different amplitudes in the NR3 and R7 data families. 
    The amplitudes for the NR3 plot are $(7.0, 7.22, 8.0)\times 10^{8}$, and for R7 $(8.6, 8.808, 9.6)\times 10^{8}$ from the bottom to top.
    To make the plots easier to read, only a subset of the solutions shown in figure \ref{f.rpmin.Eb_positive} are shown here.
  }
\end{figure}

Previous studies in spherical symmetry \cite{Andreasson2006,Akbarian2014} found support for the critical solution being an unstable steady state. 
The time derivatives of the metric fields provide some indication as to the stationarity of the solution.
In figure \ref{f.dts.Eb_positive} we show $\ln \| s_{,t }\|_{L^2}$, which is computed from the right-hand side of \eref{e:dts}. 
The plot shows that for near-critical data this quantity approaches a small value and remains small for a longer period of time the closer the data is to being critical. 
We also note that near-critical solutions spend an increased amount of time at the large radius obtained immediately after the first expansion phase.
This is most clearly seen in movies of the spatial density (see 
\href{https://youtu.be/4hmknanCf2A}{movie 2}
for a solution which recollapses, and 
\href{https://youtu.be/QfK5dtkhNpk}{movie 3}  
for a solution that eventually leaves the domain),
but can also be seen in the plots of $\min r_\mathrm{p}$ in figure \ref{f.rpmin.Eb_positive}.
The flattening at large values of $\min r_\mathrm{p}$ in both plots is a numerical artefact due to the matter leaking out of the domain. 
For these solutions, the number of particles and total mass rapidly go to zero.

\subsubsection{Families where the critical solution has 
negative binding energy
}
\label{s.results.Eb_negative}

In this class we consider three initial data families (bottom three rows of table \ref{t.initial_data}).

\begin{figure}[h]
  \centering
  \begin{subfigure}[b]{0.49\linewidth}
    \centering
    \includegraphics[width=\textwidth]{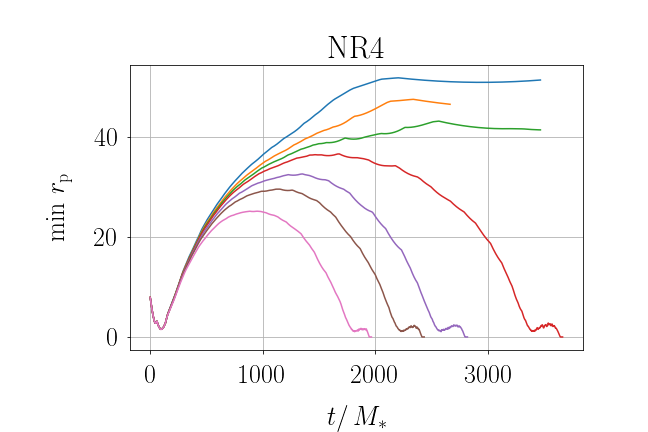}
  \end{subfigure} 
  \begin{subfigure}[b]{0.49\linewidth}
    \centering
    \includegraphics[width=\textwidth]{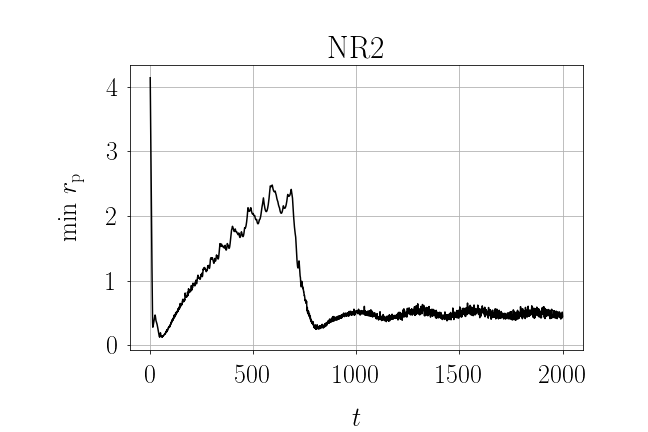}
  \end{subfigure}
  \caption{ \label{f.rpmin.Eb_negative}
    Spatial location of the matter for a few near-critical evolutions of family NR4 (amplitudes $A = (9.75, 9.775, 9.788, 9.8, 9.9) \times 10^{8}$ from the top to the bottom curve) and for a subcritical evolution of family NR2 (amplitude $A=0.071$).
    Shown is the minimum value of the radial coordinate $r_{p}$ of the particles.
  (For family NR2 we have not rescaled the $t$ axis by the mass because it changes drastically in the course of the evolution, cf. figure \ref{f.nrf2}.)
  }
\end{figure}

\begin{figure}[h]
  \centering
  \begin{subfigure}[b]{0.49\linewidth}
    \centering
    \includegraphics[width=\textwidth]{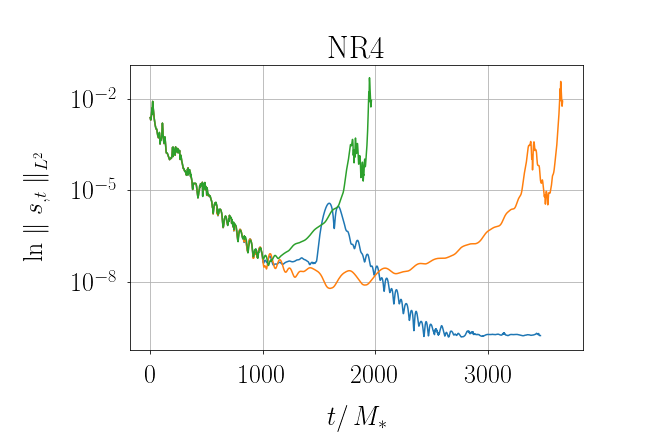}
  \end{subfigure} 
  \begin{subfigure}[b]{0.49\linewidth}
    \centering
    \includegraphics[width=\textwidth]{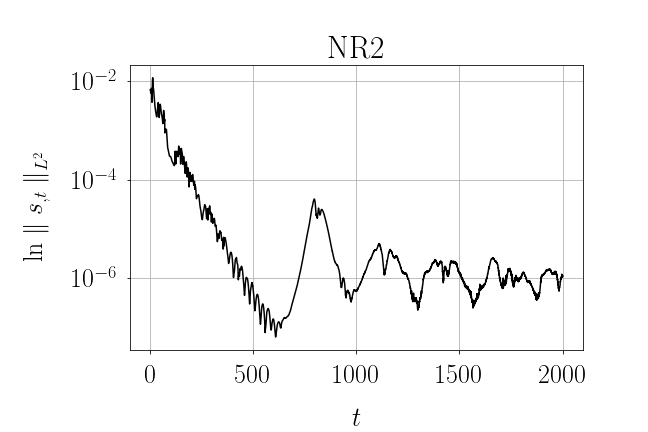}
  \end{subfigure}
  \caption{ \label{f.dts.Eb_negative}
  $L^2$-norm of the log of the right-hand side of the $s$-evolution equation \eref{e:dts} as a function time rescaled by the mass of the critical solution,
  shown for several different amplitudes in the NR4 and NR2 data families. 
  The amplitudes are $A = (9.75, 9.8, 9.9) \times 10^{8}$ from the bottom to top for NR4 and $A = 0.071$ for NR2 (cf. figure \ref{f.rpmin.Eb_negative}.)
  For family NR2 we have not rescaled the $t$ axis by the mass because it changes drastically in the course of the evolution, cf. figure \ref{f.nrf2}.
  }
\end{figure}

Family NR4 is similar to the families in section \ref{s.results.Eb_positive} except that the binding energy is now negative by shifting the support of $v_1$.
All members of this family undergo an initial collapse followed by an expansion (figure \ref{f.rpmin.Eb_negative}).
In contrast to the positive binding energy case though, supercritical solutions then collapse without further oscillations even close to the critical amplitude.
As the critical amplitude is approached from above, the lifetime of the solution increases further and further and the critical solution appears to be stationary
(\href{https://youtu.be/IYGlp9zdwTY}{movie 4}).
This is also apparent in metric fields such as $s$, the norm of whose time derivative is shown in figure \ref{f.dts.Eb_negative}: the closer to criticality, the longer $\| s_{,t}\|_{L^2}$ stays on a small level.
Subcritical solutions eventually leave the domain, whose boundary is at $r_{\max} = 60$. 
  
\begin{figure}[h]
  \centering
  \begin{subfigure}[b]{0.49\linewidth}
    \centering
    \includegraphics[width=\textwidth]{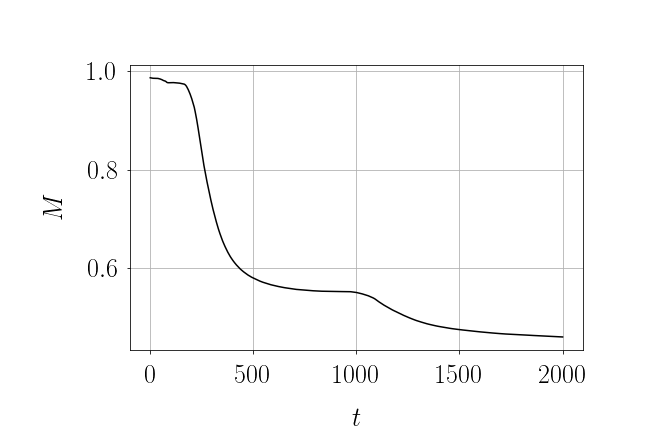}
  \end{subfigure} 
  \begin{subfigure}[b]{0.49\linewidth}
    \centering
    \includegraphics[width=\textwidth]{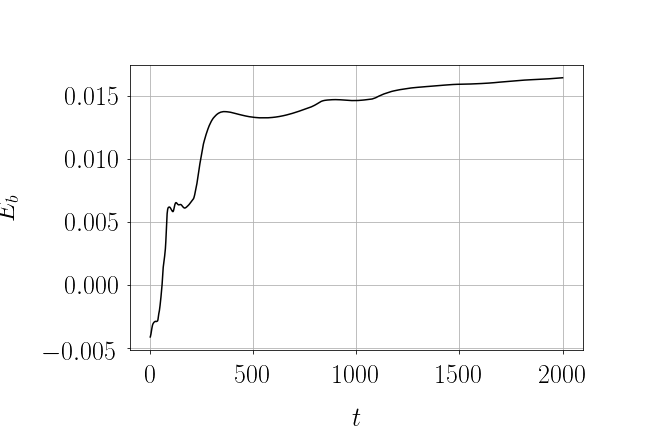}
  \end{subfigure}
  \caption{ \label{f.nrf2}
    Further illustrations of the dynamics of a subcritical evolution of family NR2 (amplitude $A=0.071$): ADM mass $M$ and binding energy $E_b$ as functions of time, measured on the finite computational domain.
  }
\end{figure}

The other two families NR2 and R4 are very different as they have much larger velocities.
R4 has been obtained from NR2 by slightly shifting the support of $v_3$ in order to obtain nonzero net angular momentum.
The behaviour of these two families is very similar.
An initial implosion is followed by a violent explosion.
Particles with high velocities soon leave the domain, leading to a significant decrease of the mass (figure \ref{f.nrf2}).
Thereafter the solution changes character:
the remaining matter that still hangs together actually has \emph{positive} binding energy.
Supercritical evolutions collapse after a series of damped oscillations as for the families with $E_b^* > 0$.
Subcritical runs appear to ring down to a stationary solution (figures \ref{f.rpmin.Eb_negative} and \ref{f.dts.Eb_negative}, and 
\href{https://youtu.be/4agWWlJimWY}{movie 5}).

\begin{figure}[h]
  \centering
  \begin{subfigure}[b]{0.5\linewidth}
    \centering
    \includegraphics[width=\textwidth]{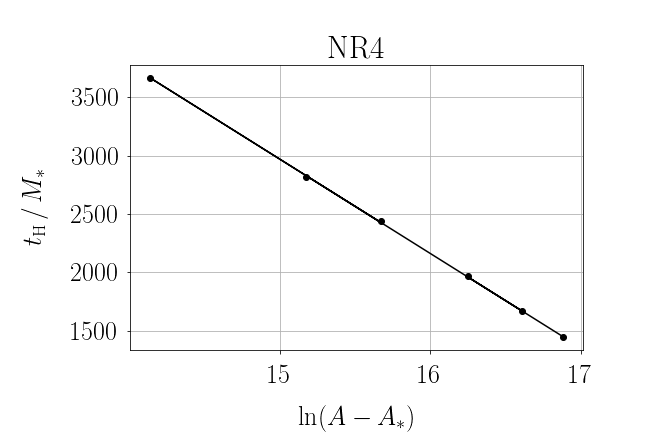}
    \label{f:tscaling_nrf4}
  \end{subfigure}\\
  \begin{subfigure}[b]{0.5\linewidth}
    \centering
    \includegraphics[width=\textwidth]{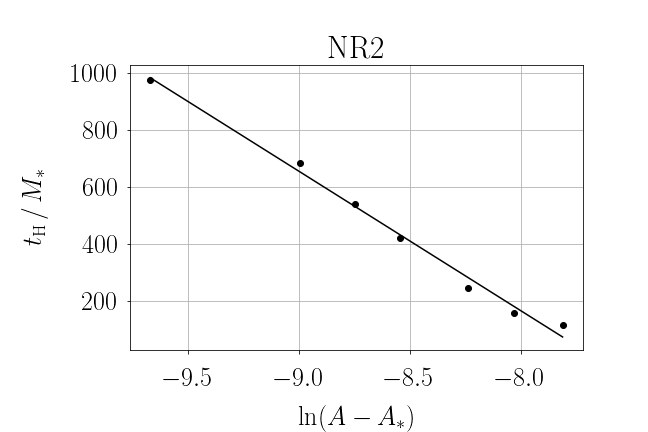}
    \label{f:tscaling_nrf2}
  \end{subfigure}\qquad
  \begin{subfigure}[b]{0.5\linewidth}
    \centering
    \includegraphics[width=\textwidth]{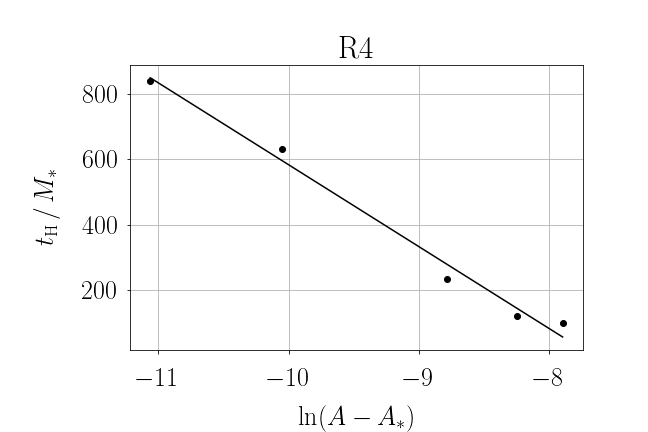}
    \label{f:tscaling_rf4}
  \end{subfigure}
  \caption{ \label{f.tscaling.Eb_negative}
    Lifetime scaling for initial data families NR4, NR2 and R4 with negative binding energy.
    Shown is the time $t_\mathrm{H}$ when the apparent horizon forms (rescaled by the ADM mass $M_*$ of near-critical data) for several amplitudes $A$, as well as the best fit of the form \eref{e:tscaling}.
    The scaling exponents are found in table \ref{t.near_critical}.}
\end{figure}

For all three families with $E_b^* < 0$ we observe lifetime scaling in supercritical evolutions (figure \ref{f.tscaling.Eb_negative}).
The corresponding scaling exponents can be found in table \ref{t.near_critical}.

\subsubsection{Families with angular momentum exceeding the square of the mass}
\label{s.results.super_rotating}

We have also constructed initial data with $|J|>M^2$.
Such data cannot collapse to a black hole immediately because the cosmic censorship conjecture implies
$|J_\mathrm{H}|<M_\mathrm{H}^2$.
Interestingly, for supercritical solutions we observe similar dynamics as in the $|J|<M^2$ cases discussed above.
In positive binding energy supercritical evolutions, the solution performs several damped oscillations similar to the case described in section \ref{s.results.Eb_positive}.
At each oscillation, a fraction of the particles are ejected, which leads to a decrease of the angular momentum of the remnant matter configuration 
(\href{https://youtu.be/ACtlHJ4bs9Q}{movie 6}).
Eventually the matter collapses and forms a black hole. 
We have some indications that this happens when the remnant has angular momentum $|J|<M^2$ (when restricting the integration domain to the support of the remnant), as one would expect.
Similarly, in negative binding energy cases, supercritical evolutions initially collapse, then expand blowing off a portion of the matter. The remnant recollapses and, once $|J|<M^2$, forms a black hole without damped oscillations (as in section \ref{s.results.Eb_negative}).
However, currently our evolutions of these data suffer from considerable numerical violations of angular momentum conservation so that we are unable to study this process in more detail.
Therefore we defer this interesting question to future work.


\section{Summary and discussion}
\label{s:conclusion}

We numerically investigate solutions launched from initial data of the form \eref{e:initialdata} for the Einstein--Vlasov system in axisymmetry. 
By tuning the amplitude of these data we probe the critical regime near the threshold of black hole formation in settings with and without net angular momentum.
This is the first study of critical behaviour beyond spherical symmetry for the Einstein--Vlasov system.
Our numerical experiments indicate that the sign of the binding energy determines the dynamics of marginally supercritical solutions. 
In the case that such solutions have positive binding energy, the formation of a black hole is preceded by a series of damped oscillations, whose initial amplitude and period increase as the critical parameter is approached from above. 
In contrast, the case in which marginally supercritical solutions have negative binding energy displays no such oscillations and solutions simply collapse after an expansion phase. 

Lifetime scaling of the marginally supercritical solutions is computed from the time to horizon formation, and we find excellent agreement with a scaling law of the form \eref{e:tscaling}.    
Our results (table \ref{t.near_critical}) indicate no universality of the scaling exponents.
This result is consistent with the conclusions of \cite{Akbarian2014} for massive particles in spherical symmetry.

Studies of critical collapse in spherical symmetry have found type I critical behaviour in which the critical solutions are static. 
It is natural in the present setting to expect the critical solutions to be stationary. 
Indeed, we find some support for this in the metric fields and energy density of the matter. 

We find no indication that the stationary critical solutions are unique.
This also is consistent with the situation in spherical symmetry.
As shown in \cite{Andreasson2006} there are infinite-dimensional families of unstable static solutions to the spherically symmetric Einstein--Vlasov system whose members can each be regarded as a critical solution. 
Thus, there is no universal critical solution in that setting.
The situation is similar in axisymmetry; a number of infinite-dimensional families of stationary solutions have been numerically constructed in \cite{Ames2016,Shapiro1993a,Shapiro1993b}.
In follow-up work we plan to use perturbative studies to investigate the viability of such solutions (those on the unstable branch) as critical solutions.

As presented above we find critical solutions with positive binding energy, and evidence that such critical solutions are stationary.
This is fully consistent with the picture of static solutions in spherical symmetry.
In that setting, there is strong numerical evidence \cite{Andreasson2006,Gunther2020} that steady states become unstable at the peak in the binding energy, and that, in contrast to claims made in \cite{Akbarian2014}, there are solutions on the unstable branch with positive binding energy, which are viable critical solutions.
In particular, in \cite{Andreasson2006,Gunther2020} it is shown that collapse-promoting perturbations of such solutions lead to black hole formation, while dispersion-promoting perturbations lead to nonsingular spacetimes. 

Our evolutions of subcritical data with positive binding energy are observed to leave the computational domain.
As noted above, due to proposition \ref{prop.complete_dispersion}, such solutions do not in fact disperse.
This observation can be compared with the dispersion-promoting perturbations of unstable steady states in the spherically symmetric setting.
G\"{u}nther et al. \cite{Gunther2020} find that in cases such perturbed states expand significantly in radius before reimploding and oscillating about a new state. 
It is possible that our computational domain is not large enough to capture such an expansion, although we have tried to capture such dynamics. 
Further clarifying the fate of these solutions will likely have to await an improved version of the code.

Finally, we comment on our classification of sub- and supercritical solutions.
We classify as supercritical solutions those which we evolve all the way to black hole formation. 
The estimated critical amplitudes in table \ref{t.near_critical} are obtained from such evolutions. 
Similarly, runs that eventually leave the domain are classified as subcritical. 
There are however initial data which we are unable to definitively classify as either sub- or supercritical, for example the middle traces in figure \ref{f.rpmin.Eb_positive}. 
The reason is that at late times ($t \gtrsim 7500 M$)
an instability grows in corners of the domain, the mass grows, and the validity of the simulation breaks down. 
We emphasize that prior to this time, the mass is conserved at the one percent level and the normalized momentum constraints remain small (quantities defined in \eref{e.normalized_residuals} are below the one percent level) and well controlled. 
However, this instability prohibits us from making a closer determination of the critical solutions.
This is particularly evident in figure \ref{f.rpmin.Eb_positive}, where solutions that are subcritical according to the lifetime scaling appear to recollapse.
These solutions may in fact be supercritical, were we able to follow them all the way to black hole formation; or, they could evolve via an oscillatory phase to a non-singular end state.
We leave it to future work to definitively classify such solutions, and more closely identify the critical solutions. 

\section*{Acknowledgments}

This research was partially supported through the programme ``Research in Pairs" by the Mathematisches Forschungsinstitut Oberwolfach in 2020, 
and by the Erwin Schr\"{o}dinger International Institute for Mathematics during the 2017 workshop on Geometric Transport Equations in General Relativity.
EA thanks the Knut and Alice Wallenberg foundation for  support during a portion of this work. 

Computations were performed on resources at Chalmers Centre for Computational Science and Engineering (C3SE) provided by the Swedish National Infrastructure for Computing (SNIC), and on the Minerva computing cluster hosted by the Max Planck Institute for Gravitational Physics in Potsdam, Germany.


\appendix

\section{Further details on the geometry equations}
\label{s:appendix.geodetails}


\subsection{Solving the constraints initially}
\label{s:appendix.geodetails.iniconstr}

At the initial time all four constraint equations
\eref{e:momconsr}--\eref{e:hamcons} need to be solved.
Unfortunately the $\psi$-scaling of the matter source terms in \eref{eq:ScaledMatter} determined by the requirement that those be $\psi$-independent (cf. \eref{e:tautildeframe}--\eref{e:Sddtildeframe}) means that the momentum constraints \eref{e:momconsr}--\eref{e:gercons} no longer decouple from the Hamiltonian constraint.
This suggests the following procedure.

For the initial data, instead of $\tilde f = \psi^6 e^{2rs} f$ we freely specify
$\hat f := \psi^2 \tilde f = \psi^8 e^{2rs} f$
and compute the corresponding source terms $\hat \rho_H := \psi^2 \tilde \rho_H$, 
$\hat J_A := \psi^2 \tilde J_A$ and $\hat J^\phi := \psi^2 \tilde J^\phi$.
Instead of $\tBp$, we consider $\hat B^\phi := \psi^{-3} \tBp$ as free data.
In order to solve the angular momentum constraint \eref{e:gercons}, we introduce the following decomposition of the vector $\tE^A$:
\begin{equation} \label{e:initwist}
  \tE^r = \psi^{-3} \rme^{-2rs} (\mu_{,r} + \nu_{,z}), \quad
  \tE^z = \psi^{-3} \rme^{-2rs} (\mu_{,z} - \nu_{,r}- 3 r^{-1} \nu).
\end{equation}
Similarly, we express the extrinsic curvature variables $\tU$ and $\tX$
in terms of a vector potential $V^A$ \cite{Rinne2008}, compare \eref{e:shiftr}--\eref{e:shiftz}:
\begin{equation} \label{e:iniUX}
  \tU = V^z{}_{,z} - V^r{}_{,r}, \quad \tX = \half(V^r{}_{,z} + V^z{}_{,r}).
\end{equation}
The fields $s$, $\tW$, $\nu$ and $\hat B^\phi$ constitute the free initial data.
 
With these choices the initial constraints are solved in the
following hierarchical order, where \eref{e:momconsrini} and \eref{e:momconszini} form a coupled linear system: 
\begin{eqnarray}
  \label{e:gerconsini}
  \fl 0 = \mu_{,rr} + 3 r^{-1} \mu_{,r} + \mu_{,zz}
     - 2 \kappa \hat J^\phi,\\
  \label{e:momconsrini}
  \fl 0 = \tfrac{2}{3} V^r{}_{,rr} + V^r{}_{,zz} + \third V^z{}_{,rz}
    + 2 R_r (V^r{}_{,r} - V^z{}_{,z}) + 2 R_z (V^r{}_{,z} + V^z{}_{,r}) \nonumber\\
    + \tfrac{2}{3} r \tW_{,r} + \tfrac{8}{3} \tW 
    + r^2 \hat B^\phi (\nu_{,r} - \mu_{,z} + 3r^{-1}\nu) - 2 \kappa e^{-rs} \hat J_r, \\
  \label{e:momconszini}
  \fl 0 = -\third V^r{}_{,rz} + V^z{}_{,rr} + \tfrac{4}{3} V^z{}_{,zz}
   + (2 R_r + r^{-1}) (V^r{}_{,z} + V^z{}_{,r}) + 2 R_z (V^z{}_{,z} - V^r{}_{,r}) \nonumber\\
   + \tfrac{2}{3} r \tW_{,z} + r^2 \hat B^\phi (\mu_{,r} + \nu_{,z})
   - 2 \kappa e^{-rs} \hat J_z,\\
  \label{e:hamconsini}
  \fl 0 = \psi_{,rr} + \psi_{,zz} + r^{-1} \psi_{,r} + \fourth \psi [ r s_{,rr} +
    r s_{,zz} + 2 s_{,r} ] \nonumber\\ + \fourth \psi^{-7} \exps \left[
    \third (V^z{}_{,z} - V^r{}_{,r} + \half r \tW)^2 + \fourth (r \tW)^2 
    + \fourth (V^r{}_{,z} + V^z{}_{,r})^2 \right]
  \nonumber\\
  + \tfrac{1}{16} r^2 \psi^{-7}\left[(\mu_{,r} + \nu_{,z})^2
    + (\mu_{,z} - \nu_{,r}- 3 r^{-1} \nu)^2\right]  \nonumber\\
  + \tfrac{1}{16} r^2 \psi e^{2rs} (\hat B^\phi)^2
  + \fourth \kappa \psi^{-3} \hat \rho_H .
\end{eqnarray} 

Finally, now that we have $\psi$, we rescale the Vlasov distribution function $\tilde f := \psi^{-2} \hat f$ and $\tBp := \psi^3 \hat B^\phi$, compute the vector $\tE^A$ from \eref{e:initwist}, and $\tU$ and $\tX$ from \eref{e:iniUX}.


\subsection{Boundary conditions}
\label{s:appendix.geodetails.bcs}

All evolved variables are either even functions of $r$ with nonzero value at $r=0$, or odd functions of $r$ with nonzero derivative at $r=0$:
\[ \begin{array}{ll}
  \textrm{even:} & \psi, \tU, \tE^z, \alpha, \beta^z, \mu, P, V^z
  \\
  \textrm{odd:} & s, \tW, \tX, \tE^r, \tBp, \beta^r, \nu, Q, V^r
\end{array} \] 
We use this parity information to adapt our centred finite-difference stencils near the origin so that the variables are only evaluated at grid points with $r>0$.

We turn now to the boundary conditions at the outer boundary of the computational domain, i.e. at $r=r_{\max}$, $z=z_{\max}$ and $z=z_{\min}$.
These follow from demanding that spacetime be asymptotically flat, which implies that the components of the metric approach their Minkowski values as $R\to\infty$, where here and in the following $R = \sqrt{r^2 + z^2}$ and $\theta = \tan^{-1}(r/z)$ are spherical polar coordinates.

More precisely, for a variable $u$ that obeys an elliptic equation we assume that
\begin{equation} \label{e:ellipticfalloff}
  u \sim u_0 + f(\theta)/R^p 
\end{equation}  
as $R\to\infty$ with some (unknown) function $f(\theta)$, a constant $u_0$ and a positive integer $p$.
The powers $p$ can be inferred from an analysis of the asymptotic behaviour of solutions to the linearised Einstein equations with spherical harmonic indices $\ell=2$ and $\ell=3$ \cite{Rinne2008a}.
The values of $u_0$ and $p$ will be given below for each of the evolved variables.
By taking an $R$-derivative of \eref{e:ellipticfalloff}, we obtain a Robin condition
\begin{equation} \label{e:robinbc}
  0 \doteq \partial_R [R^p (u-u_0)].
\end{equation}
This is the condition we impose at the outer boundary of the computational domain ($\doteq$ denoting equality at the outer boundary).

For a variables $u$ that obeys an evolution equation, we assume
\begin{equation} \label{e:evolutionfalloff}
  u \sim u_0 + f(t-R, \theta)/R^p,
\end{equation}
which reflects the fact that the fields are outgoing to leading order in $1/R$.
By taking derivatives of \ref{e:evolutionfalloff}, we obtain a Sommerfeld condition
\begin{equation} \label{e:sommerfeldbc}
  0 \doteq (\partial_t + \partial_R) [R^p (u-u_0)].
\end{equation}

We now provide the values of the constants $p$ and $u_0$ for each of the evolved variables.
For the initial constraints \eref{e:gerconsini}--\eref{e:hamconsini}, we use Robin boundary conditions \eref{e:robinbc} on $\mu$, $V^r$, $V^z$ and $\psi$, with $\mu_0 = V^r_0 = V^z_0 = 0$, $\psi_0 = 1$, and $p=1$ in all cases.

When solving the Hamiltonian constraint \eref{e:hamcons} during the evolution, we impose a Dirichlet condition on $\psi$ with data taken from the evolution equation \eref{e:dtpsi} for $\psi$ at the outer boundary;
this ensures that the constraint is compatible with the evolution equation.
Robin conditions \eref{e:robinbc} are used for $\alpha$ and the shift potentials $P$ and $Q$, with $\alpha_0=1$, $P_0=Q_0=0$, and $p=1$ in all cases.

We use Sommerfeld conditions \eref{e:sommerfeldbc} for the evolution equations \eref{e:dtpsi}--\eref{e:dtBp_} on $\psi, s, \tW, \tU, \tX, \tE^r, \tE^z$ and $\tBp$, with $\psi_0 = 1$, $s_0 = \tW_0 = \tU_0 = \tX_0 = \tE^r_0 = \tE^z_0 = \tBp_0 = 0$, $p=2$ for $s, \tW$ and $p=1$ for the remaining variables.

We stress that since no exact boundary conditions are known for asymptotically flat solutions to Einstein's field equations at a finite distance, imposing boundary conditions at the outer boundary of our finite computational domain that are correct at the leading order in $1/R$ is the best we can do.
This does introduce small reflections off the outer boundary, including violations of the momentum constraints, but they remain under control during the evolution.
Similar boundary conditions have been used before e.g. in \cite{Choptuik2003a,Choptuik2003b,Rinne2008}.


\section{Diagnostics}
\label{s:appendix.diagnostics}


\subsection{Apparent horizons}
\label{s:appendix.diagnostics.horizons}

An apparent horizon or marginally outer trapped surface is the outermost 2-surface in a spatial slice whose outgoing null expansion vanishes.
Following section 9.3 of \cite{RinnePhD}, this condition reads in (2+1)+1 form
\begin{equation}
  \label{e:aheq211}
  s^A{}_{\parallel A} + \lambda^{-1} s^A \lambda_{,A} + \chi_{AB} s^A s^B
  - \chi - K_\phi^\phi = 0,
\end{equation}
where $s^A$ is the unit outward normal to the curve that represents the horizon
in the spatial 2-slice of $\mathcal{N}$.
Parametrising the coordinates $x^A$ of this curve by $\tau$, the normal
takes the form
\begin{equation}
  \label{e:ahnormal}  
  s^A = N H^{AB} \varepsilon_{BC} \frac{\rmd x^C}{\rmd\tau},\quad
  N := \left( H_{AB} \frac{\rmd x^A}{\rmd\tau}\frac{\rmd x^B}{\rmd\tau}
  \right)^{-1/2}
\end{equation}
and one can show \cite{Nakamura1987}
\begin{equation}
  s^A{}_{\parallel A} = -N^3 \varepsilon_{AB} \left( \frac{\rmd^2 x^A}{\rmd\tau^2}
    \frac{\rmd x^B}{\rmd\tau} + \Gamma^A{}_{CD}
    \frac{\rmd x^B}{\rmd\tau}\frac{\rmd x^C}{\rmd\tau}\frac{\rmd x^D}{\rmd\tau}
    \right).
\end{equation}
In practice, we choose the curve parameter $\tau$ to be the spherical polar
angle $\theta$, so the coordinates of the curve are given by
\begin{equation}
  x^1 = r = R(\theta) \sin\theta, \quad x^2 = z = R(\theta) \cos\theta,
\end{equation}
where $R(\theta)$ is the function to be determined.
With our choice of gauge and variables \eref{e:aheq211} reads
\begin{equation}
  \label{e:aheq}
  \fl R''(\theta) = a_0 + (4 P_r + R_r) a_1 + (4 P_z + R_z) a_2
  + \psi^{-4} e^{rs} (\tW a_3 + \tU a_4 + \tX a_5),
\end{equation}
where, with $\ell := \sqrt{R^2 + R'^2}$,
\begin{eqnarray}
  a_0 &=& 2 R + \frac{3 R'^2}{R} - \frac{\ell^2 R'}{R^2} \cot\theta ,\\
  a_1 &=& \ell^2 \sin\theta - \frac{\ell^2 R'}{R} \cos\theta,\\
  a_2 &=& \frac{\ell^2 R'}{R} \sin\theta + \ell^2 \cos\theta,\\
  a_3 &=& \frac{\ell^3}{3} \sin\theta,\\
  a_4 &=& 2 R' \ell \cos\theta \sin\theta
  + \frac{\ell(R^2 - R'^2)}{R} \cos^2\theta - \frac{\ell(R^2 - 2 R'^2)}{3 R},\\
  a_5 &=& \frac{2 \ell (R^2 - R'^2)}{R} \cos\theta\sin\theta
  - 2 R' \ell (2 \cos^2\theta - 1).
\end{eqnarray}

The irreducible mass of the apparent horizon is 
$M_\mathrm{irr} = \sqrt{A_\mathrm{H}/(16\pi)}$, where
the horizon area is given by
\begin{equation}
  A_\mathrm{H} = 2\pi \int_\mathrm{H} \lambda \, \rmd s, \qquad
  \rmd s^2 = H_{AB} \frac{\rmd x^A}{\rmd\theta}\frac{\rmd x^B}{\rmd\theta}
  \rmd\theta^2.
\end{equation}
With our choice of gauge and variables we obtain
\begin{equation}
  A_\mathrm{H} = 2\pi \int_{0}^{\pi} \psi^4 \rme^{rs} \, \ell R \sin\theta \, \rmd\theta.
\end{equation}

The angular momentum of the horizon can be determined
using the angular momentum constraint \eref{e:gercons}.
This equation can be written in covariant form as
\begin{equation}
  (2\kappa)^{-1} (\lambda^3 E^A)_{\parallel A} = \lambda^3 J^\phi,
\end{equation}
where $\parallel$ denotes the covariant derivative of the 2-metric $H$.
Integrating this equation and using Gauss' law, the boundary term at the horizon is the angular momentum of the black hole:
\begin{equation}
  \label{e:Jhorizon} \label{e:J_H}
  J_\mathrm{H} = -(2\kappa)^{-1} \int_\mathrm{H} \lambda^3 E^A s_A \rmd s
  = -\frac{1}{8} \int_0^\pi \lambda^3 \varepsilon_{AB} E^A
  \frac{\rmd x^B}{\rmd\theta} \rmd\theta.
\end{equation}
With our choice of coordinates and variables, \eref{e:Jhorizon} reads
\begin{equation}
  \label{e:Jhorizon2}
  \fl J_\mathrm{H} = \frac{1}{8} \int_0^{\pi}
  r^3 \psi^3 e^{2rs} \big[ \tE^r (R\sin\theta - R'\cos\theta )
  + \tE^z (R\cos\theta + R'\sin\theta)\big]
  \rmd\theta.
\end{equation}

Finally, the black hole mass $M_\mathrm{H}$ is computed from the
irreducible mass $M_\mathrm{irr}$ and the horizon angular momentum
$J_\mathrm{H}$ via
\begin{equation} \label{e:M_H}
  M_\mathrm{H} = (2M_\mathrm{irr})^{-1} \sqrt{4 M_\mathrm{irr}^4 + J_\mathrm{H}^2}.
\end{equation}
This is only strictly valid for a stationary black hole.


\subsection{Conserved quantities}
\label{s:appendix.diagnostics.conserved}

In this subsection we present expressions for important conserved quantities of asymptotically flat axisymmetric spacetimes with Vlasov matter: the ADM mass, rest mass and angular momentum.

Dain \cite{Dain2008} derived a positive definite integral expression for the ADM mass of asymptotically flat axisymmetric spacetimes using the same quasi-isotropic gauge as we do. 
Starting from equation (68) in his paper, translating to our variables and using the Hamiltonian constraint \eref{e:hamcons}, we obtain the following manifestly non-negative expression:
\begin{eqnarray}
  \label{e:DainMass}
    \fl M = \int_{-\infty}^\infty \rmd z \int_0^\infty \Big\{
    \psi^{-2} (\psi_{,r}^2 + \psi_{,z}^2)
    + \fourth \psi^{-8} e^{2rs} \left[ \third (\tU + \half r \tW)^2
    + \fourth (r\tW)^2 + \tX^2 \right] \nonumber \\
    + \tfrac{1}{16} r^2 \psi^{-6} e^{2rs} \left[ {\tBp}^2
    + \psi^4 \exps \left( {\tE^r}^2 + {\tE^z}^2 \right) \right]
    + \fourth \kappa \psi^{-2} \tilde \rho_H \Big\} r \, \rmd r.
\end{eqnarray}
This mass $M$ is a conserved quantity.

The rest mass can be derived from the current density
\begin{equation}
  N^{\alpha} = \int f p^{\alpha}\frac{1}{\sqrt{|g|}}\frac{\rmd p_1\rmd p_2 \rmd p_3}{p^0},
\end{equation}
which is divergence-free:
\begin{equation}
  \nabla_\alpha N^\alpha = 0.
\end{equation}  
Integrating this over a spacetime region bounded by two constant-time slices $\Sigma_t$ with unit normal $n$ and using Gauss' law, we obtain the conserved rest mass
\begin{equation}
  \label{e:RestMass}
  \fl M_0 = 2\pi \int_{\Sigma_{t}} \lambda \sqrt{H} n_\alpha N^\alpha \, \rmd r \, \rmd z
    = 2\pi \int_{\Sigma_{t}}  \left(  \int_{\mathbb R^3} \tilde{f}(r, z, v) \, \rmd^3 v \right) r \, \rmd r \, \rmd z
\end{equation}
in terms of our frame variables \eref{e:MomentumFrame}, where $\tilde f$ is the rescaled distribution function \eref{eq:RescaledDensity}.

The existence of the Killing vector $\xi$ related to the axisymmetry immediately gives rise to a conserved angular momentum, which can be defined by the Komar integral 
\begin{equation}
  J = (2\kappa)^{-1} \oint_{S^2} \rmd S_{\mu\nu} \nabla^\mu \xi^\nu,
\end{equation}
where $S^2$ is a 2-sphere within a spatial slice $\Sigma_t$ outside the matter support and $S_{\mu\nu}$ is its area element.
Assuming that spacetime has a regular centre, we can convert this expression to a volume integral.
Using Einstein's equations, we obtain
\begin{equation}
  \label{e:J}
  \fl J = 2 \pi \int_{\Sigma_t} \lambda^3 \sqrt{H} J^\phi \, \rmd r \, \rmd z
    = 2\pi \int_{\Sigma_{t}}  \left(  \int_{\mathbb R^3} L \tilde{f}(r, z, v) \, \rmd^3 v \right) r \, \rmd r \, \rmd z,
\end{equation}
which is clearly conserved because $L$ is.
Note however that in terms of our frame variables, $L = r \psi^2 v_3$, and $v_3$ obeys the evolution equation \eref{e:dv3dt}.



\section*{References}

\bibliographystyle{iopart-num-long}
\bibliography{references}

\end{document}